\documentclass[bibyear]{aa} 

\usepackage{txfonts}
\usepackage{graphicx}   
\usepackage{amsmath}    
\usepackage[utf8]{inputenc} 
\usepackage{array} 
\usepackage{multicol,rotating} 
\usepackage{longtable} 
\usepackage[section]{placeins}
\usepackage{lscape}

\begin{document}

   \title{Volatile exposures on the 67P/Churyumov-Gerasimenko nucleus}

   \author{S. Fornasier
          \inst{1,2}
          \and
          H. V. Hoang\inst{1,3} \and
          M. Fulle\inst{4}  \and
          E. Quirico\inst{3} \and
          M. Ciarniello\inst{5}
          }
   \institute{LESIA, Universit\'e Paris Cit\`e, Observatoire de Paris, Universit\'e PSL, CNRS, Sorbonne Universit\'e, 5 place Jules Janssen, 92195 Meudon, France
              \email{sonia.fornasier@obspm.fr}
         \and
           Institut Universitaire de France (IUF), 1 rue Descartes, 75231 Paris Cedex 05  
        \and  
        Universit\'e Grenoble Alpes, CNRS, Institut de Plan\'etologie et Astrophysique de Grenoble (IPAG), UMR 5274, Grenoble F-38041, France
        \and
        Osservatorio Astronomico, INAF, Trieste, Italy 
        \and
        Istituto di Astrofisica e Planetologia Spaziali (IAPS), Istituto Nazionale di Astrofisica (INAF), Rome, Italy
             }

   \date{Accepted for publication on Astron. Astroph. on Feb. 2023}

 
  \abstract
   {}
   {We present the most extensive catalog of exposures of volatiles on the 67P/Churyumov-Gerasimenko nucleus generated from observations acquired with the Optical, Spectroscopic, and Infrared Remote Imaging System (OSIRIS) on board the Rosetta mission. We investigate the volatile exposure distribution across the nucleus, their size distribution, and their spectral slope evolution.}
   {We analyzed medium- and high-resolution images acquired with the Narrow Angle Camera (NAC) of OSIRIS at several wavelengths in the 250--1000 nm range, investigating images from 109 different color sequences taken between August 2014 and September 2016, and covering spatial resolution from a few m/px to 0.1 m/px. To identify the icy bright spots, we adopted the following criteria: i) they should be at least 50\% brighter than the comet dark terrain; ii) they should have neutral to moderate spectral slope values in the visible range (535 -882 nm);  iii) they should be larger than 3 pixels.}
   {We   identified more than 600 volatile exposures on the comet, and we analyzed them in a homogeneous way. Bright spots are found isolated on the nucleus or grouped in  clusters, usually at the bottom of cliffs, and most of them are small,  typically a few square meters or smaller.  The isolated ones are observed in different types of morphological terrains, including smooth surfaces, on   top of boulders, or close to irregular structures. Several of them are clearly correlated with the cometary activity, being the sources of jets or appearing after an activity event.  We note a number of peculiar exposures of volatiles with negative spectral slope values in the high-resolution post-perihelion images, which we interpret as the presence of large ice
grains ($>$ 1000 $\mu$m)  or local frosts condensation. We observe a clear difference both in the spectral slope and in the area distributions of the bright spots pre- and post-perihelion, with these last having lower average spectral slope values and a smaller size, with a median surface of 0.7 m$^2$, even if the size difference is mainly due to the higher resolution achieved post-perihelion. The minimum duration of the bright spots shows three clusters:  an area-independent cluster dominated by   short-lifetime frosts; an area-independent cluster with lifetime of 0.5--2 days, probably associated with the seasonal fallout of dehydrated chunks; and an area-dependent cluster with lifetime longer than 2 days consistent with water-driven erosion of the nucleus.}
   {Even if numerous bright spots are detected, the total surface of exposed water ice is less than 50000 m$^2$, which is 0.1\% of the total 67P nucleus surface. This confirms that the surface of comet 67P is dominated by refractory dark terrains, while exposed ice occupies only a tiny fraction. High spatial resolution is mandatory to identify ice on cometary nuclei surfaces. Moreover, the abundance of volatile exposures is six times less in the small lobe than in the big lobe, adding additional evidence to the hypothesis that comet 67P is composed of two distinct bodies. The fact that the majority of the bright spots identified have a surface lower than 1 m$^2$ supports a model in  which  water ice enriched blocks (WEBs) of 0.5--1 m size should be homogeneously distributed in the cometary nucleus embedded in a refractory matrix.}

   \maketitle
%

\section{Introduction}
\begin{figure*}
\centering
\includegraphics[width=0.99\textwidth]{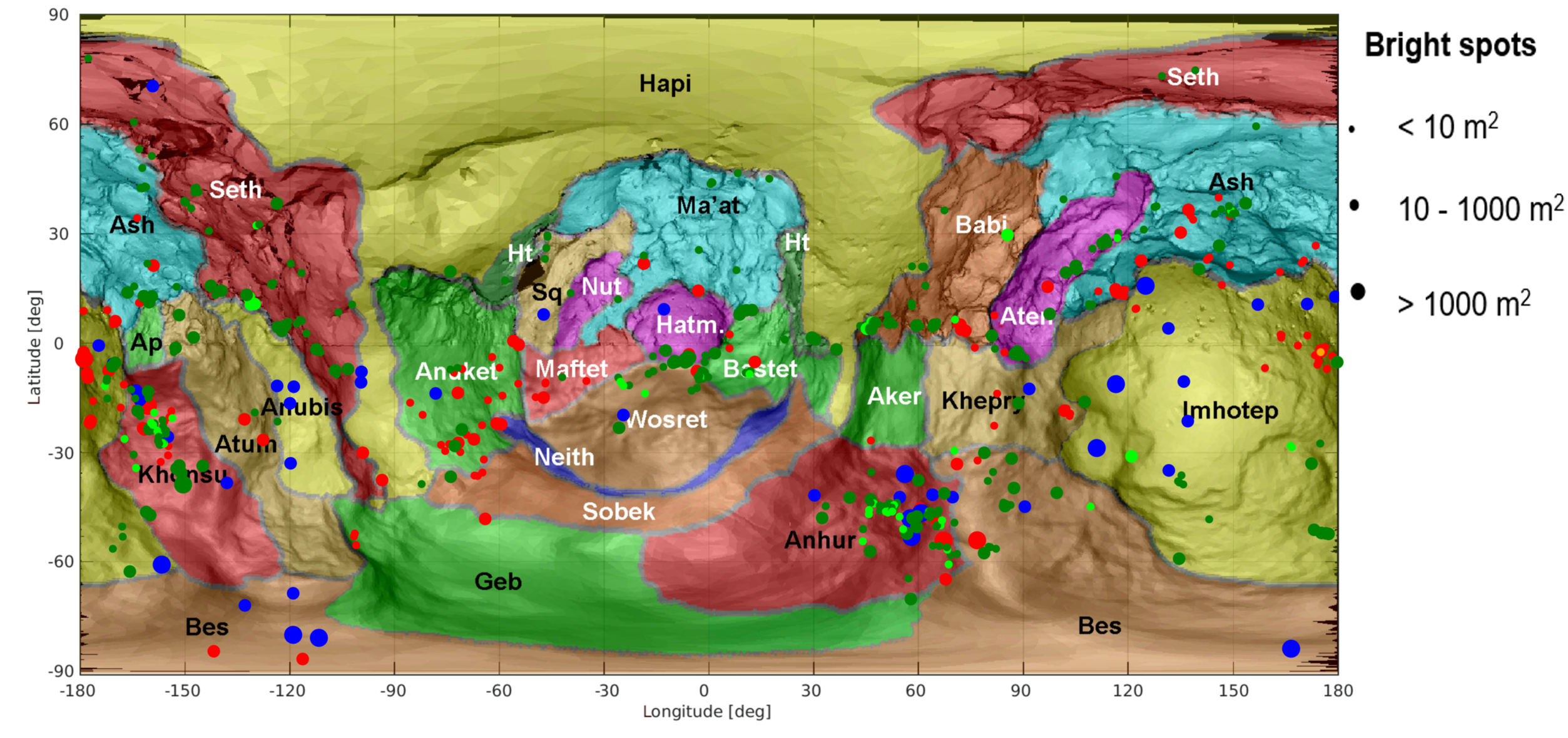}
\caption{Maps of  ice exposure on comet 67P. The  color-coding is as follows: red,  pre-perihelion (August 2014-May 2015); cyan,  perihelion (June - October 2015); green,  post-perihelion (November 2016-September 2016). The spectrally blue spots, those having negative spectral slope ($< -3\%/(100 nm)$) in the 535-882 nm range,  are shown in light green   (post-perihelion) and orange (pre-perihelion). The symbol size represents three ranges of volatile exposure area; they are not in scale compared to the nucleus surface (51.74 km$^2$ in total; Thomas et al. 2018), but are enlarged for clarity.}
\label{bsmap}
\end{figure*}

Comet 67P/Churyumov-Gerasimenko (hereafter 67P) was the main target of the Rosetta mission of the European Space Agency. Launched in 2004, Rosetta took ten years to reach the comet before orbiting around it for $\sim$ 25 months, from July 2014 to September 2016, permitting an in-depth investigation of the  67P nucleus morphology, physical properties, and composition, and of the cometary activity and the dust--gas interaction on the nucleus surface and the inner coma at different heliocentric distances. For the first time in space exploration Rosetta delivered a lander, Philae, on a cometary surface on 12 November 2014. Even if Philae rebounded from the original selected landing site,  after an adventurous trajectory and a second rebound (O'Rourke et al. 2020), it reached the Abydos site  where most of the foreseen in situ measurements were successfully achieved.        

Rosetta revealed a complex morphology of the nucleus, with different kinds of terrains (Thomas et al. 2015), including layers, boulders, cliffs, and pits, sometime active (Vincent et al. 2015). The cometary surface shows pervasive fractures ranging from millimeters (Bibring et al. 2015) to several tens of meters long produced by thermal insolation weathering (El-Maarry et al. 2015), as well as {\it goosebumps} or clod features  on  a scale of a few meters  (Sierks et al. 2015; Davidsson et al. 2016; Fornasier et al. 2021) interpreted as remnants of the original pebbles or results of fracturing processes. Twenty-six regions, named after Egyptian deities, were identified based on the surface geomorphological properties (see El-Maarry et al. 2015, 2016 for the cometary regions definition and location). The bilobate shape of the nucleus, which shows extensive layering but with different centers of gravity between the large and small lobes, is associated with a binary structure resulting  from the collision at low speed of two distinct bodies in the early Solar System (Massironi et al. 2015). The binary structure interpretation is also supported   by the different mechanical and physical properties reported for the two lobes (El-Maarry et al. 2016; Fornasier et al. 2021).

\begin{figure*}
\centering
\includegraphics[width=1.0\textwidth]{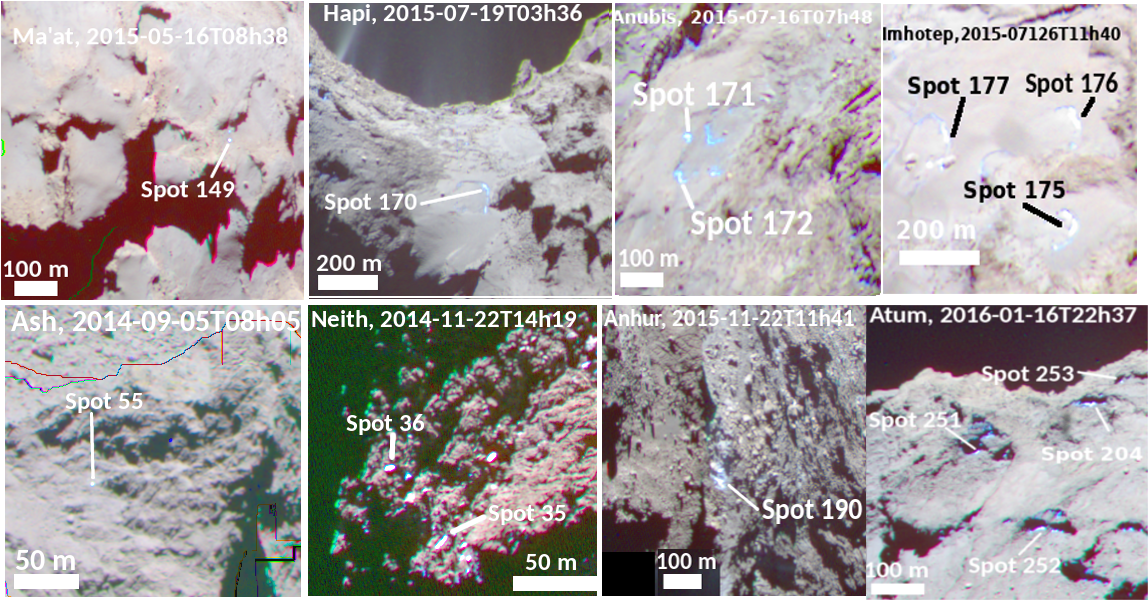}
\caption{Example of isolated bright features on smooth terrains (top) and close to irregular structures (bottom), feature types 1 and 2, respectively, following the  Deshapriya et al. (2018) classification scheme. The bright spot numbers correspond to those listed in Table~\ref{catalog}.}
\label{spots1e2}
\end{figure*}
The comet is dark with a geometric albedo of 6.5$\pm$0.2\% at 649 nm (Fornasier et al. 2015). 
The nucleus composition is dominated by refractory material mixed with opaque minerals and organics. The spectrum is red (i.e., the reflectance increases in a steep way with the wavelength)  and is characterized by a wide absorption band in the 2.8--3.6 $\mu$m region indicating the presence of a complex mixture of organics (Capaccioni et al. 2015; Quirico et al. 2016). 
The latest recalibration of the Visible, InfraRed, and Thermal Imaging Spectrometer (VIRTIS) gave evidence of different structures in the  broad band, attributed to ammonium salts (Poch et al. 2020) and aliphatic organics (Raponi et al. 2020), with a possible contribution from hydroxylated amorphous silicates to the overall absorption (Mennella et al. 2020). \\
The nucleus shows compositional heterogeneities on several spatial scales, resulting in different  spectral slopes and albedo in regional and local areas. On the dark and red average cometary terrain, exposures of volatiles stand out because they are very bright and with a   bluer spectrum (i.e., less steep).  Two volatile species were detected as exposed ice on comet 67P, mainly crystalline water ice (De Sanctis et al. 2015; Barucci et al. 2016; Filacchione et al. 2016a) and also  carbon dioxide, the latter found for the very first time exposed at a comet surface (Filacchione et al. 2016b). \\
Joint observations of the Optical, Spectroscopic, and Infrared Remote Imaging System (OSIRIS) and VIRTIS spectrometers have proven that the bright and spectrally bluer features observed with the cameras display the typical water ice bands in the infrared spectra. Based on this correlation, a number of bluer and bright features detected with OSIRIS have been attributed to exposure of water ice (Barucci et al. 2016). Pommerol et al. (2015) reported the first OSIRIS detection of volatile exposures on comet 67P with features being 5--10 times brighter than the cometary dark terrain. Desphapriya et al. (2018) generated the first catalog of volatile exposures including 57 entries. Other studies highlight the presence of bright spots associated with water ice exposures  in the northern hemisphere (Pommerol et al. 2015; Fornasier et al. 2015; Raponi et al. 2016; Barucci et al. 2016; Filacchione et al. 2016a; Lucchetti et al. 2017; La Forgia et al. 2015; De Sanctis et al. 2015; Oklay et al. 2017) and in the  southern hemisphere (Fornasier et al. 2016, 2019a, 2021; Deshapriya et al. 2016, 2018; Hasselmann et al. 2019; Hoang et al. 2020), sometimes freshly exposed on the surface after cliff collapses or outbursts (Pajola et al. 2017a; Agarwal et al. 2017; Filacchione et al. 2016a) or due to the mechanical action of Philae (O'Rourke et al. 2020). \\
The estimated water ice abundance varies from a few percent (Barucci et al. 2016; Filacchione et al. 2016a, 2016b; Raponi et al. 2016; De Sanctis et al. 2015; Ciarniello et al. 2016) to more than 20--30\% in several bright areas observed in the Imhotep, Seth, Khonsu, Bes, Anhur, and Wosret regions (Deshapriya et al. 2016, 2018; Oklay et al. 2017; Pajola et al. 2017a; Fornasier et al. 2016, 2017, 2019a, 2021; Hasselmann et al. 2019; Hoang et al. 2020), with peaks up to $\sim$ 50--80\% in few localized tiny bright spots (Oklay et al. 2017; Hoang et al. 2020; O'Rourke et al. 2020; Fornasier et al. 2021), indicating fresh exposures of volatiles.

In this paper we present the most extended catalog of exposures of volatiles of comet 67P built upon a systematic analysis of the color sequences acquired with the OSIRIS cameras. We investigate their distribution in the different cometary regions and morphological terrains, their spectral slope evolution, their size distribution, and their duration with the aim of  understanding volatile properties in comets, and of constraining cometary models.

\section{Observations and methodology}

\begin{figure*}
\centering
\includegraphics[width=1.0\textwidth]{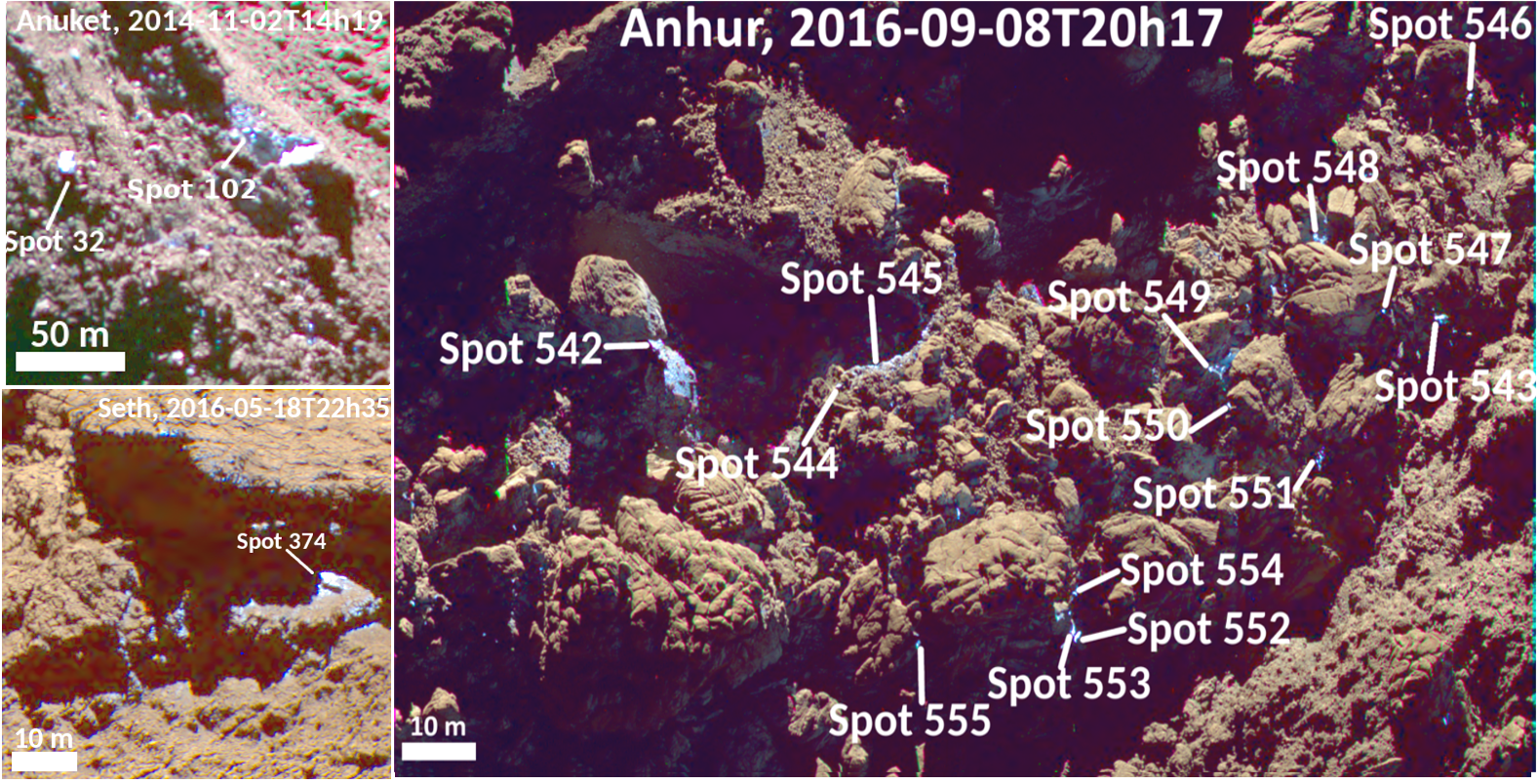}

\caption{Example of bright features resting on boulders (BS 32 in Anuket, BS 374 in Seth, and BS 542 and BS 544 in Anhur), type  3 following Deshapriya et al. (2018) classification scheme. The bright spot  numbers correspond to those listed in Table~\ref{catalog}. Several BS show blue colors, indicating a small or negative spectral slope value (see Table~\ref{catalog}).}
\label{spots3}
\end{figure*}

The analysis is based on data from the OSIRIS imaging system of the  Rosetta mission. OSIRIS included two cameras, the Narrow Angle Camera (NAC) for the high-resolution study of the nucleus, and the Wide Angle Camera (WAC) for the coma investigation (Keller et al. 2007).\\
We   analyzed medium- and high-resolution images acquired with the NAC camera with several filters in the 250--1000 nm range, investigating 109 different color sequences taken between August 2014 and September 2016, and covering a spatial resolution from a few m/px up to 0.1 m/px. We searched in the OSIRIS archive all the NAC spectrophotometric sequences pointing to the 67P nucleus and having at least three filters. Exposures of volatiles are usually brighter than the comet dark terrain, and are characterized by a neutral to moderate spectral slope in the visible range, which has been proven to be associated with a local enrichment in water ice thanks to joint observations carried out with the OSIRIS cameras and the VIRTIS spectrometer (Barucci et al. 2016; Filacchione et al. 2016a).  Thus, to determine whether   a bright feature on the surface is ice dominated and not simply brighter because of illumination conditions,   information on the reflectance value and on the spectral slope of a region of interest (ROI) are needed. This implies that some bright spots observed only in the NAC orange filter centered at 649 nm, the filter most frequently used to generate the nucleus shape model or to investigate the comet morphology, are not included in our analysis. With only one filter available we cannot determine if the higher brightness is due to a geometric effect, to the presence of a bright mineral, or to a real exposure of ice. \\
Therefore, we applied the following methodology to identify exposures of volatiles on the 67P nucleus. Bright spots exposing volatiles should be both brighter (by at least 50\%) than the comet dark terrain, and should have neutral to moderate spectral slope values in the visible range (535--882 nm). The adopted upper limit in the spectral slope value is 11\%/(100 nm), but usually bright spots have spectral slopes much lower than 8\%/(100 nm), and often close to zero. Moreover, we considered only the bright spots larger than 3 pixels. Smaller features are indeed difficult to   characterize because of residuals in the images co-registration process.\\
We used the NAC images generated by the instrument pipeline (Tubiana et al. 2015)  corrected by bias, flat field, geometric distortion, absolutely calibrated in radiance, and finally converted in radiance factor (also named $I/F$)
\begin{equation}
Radiance Factor ~(i,e,\alpha,\lambda)  =  \frac{\pi I(i,e,\alpha,\lambda)}{Fsun_{\lambda}},
\end{equation}
where I is the scattered radiance at a given incidence ($i$), emission ($e$), phase ($\alpha$) angles and wavelength ($\lambda$), and $Fsun_{\lambda}$ is the incoming irradiance of the Sun at the heliocentric distance of the comet and at a given wavelength ($\lambda$).

As done in previous studies of the 67P nucleus, the NAC images of a given sequence were first co-registered using the F22 NAC filter (centered at 649.2nm) as a reference, then corrected by the  illumination conditions using the Lommel-Seeliger disk function and the 3D stereophotoclinometric shape model of the 67P nucleus (Jorda et al. 2016), adopting the same methodology already presented in Hasselmann et al. (2019) and Fornasier et al. (2017, 2019a). \\ 
We also created RGB images with the STIFF software, which converts scientific FITS images to TIFF (Bertin 2012), mostly using the filters centered at 882 nm (R), 649 nm (G), and 480 nm (B). These RGB images are very helpful in identifying volatile exposures since they look bright and blue   compared to the dark and red cometary terrain. \\
For each bright feature, the spectral slope ($Sl$) was computed in the 535--882 nm range as:   \\
\begin{equation}
Sl = \frac{R_{882} - R_{535}}{R_{535} \times (882 ~nm - 535 ~nm)}, 
\end{equation}
where R$_{882}$ and R$_{535}$ are the radiance factors in the filters centered at 882 nm and 535 nm, respectively. \\
Details on the observing conditions are reported in Table~\ref{catalog}.

\section{Catalog of volatiles exposures}

We identified and characterized 603 bright spots (hereafter BS)  having a spectral slope much lower than the typical value of the cometary dark terrain (Table~\ref{catalog}), thus indicating local exposures of volatiles, very likely water ice on the 67P nucleus. This is the most complete catalog of volatile exposures on 67P published to date, increasing by a factor of $\sim$ 10 the number of identified bright spots on the comet compared to data already published in the literature. However, this catalog does not include the totality of the volatile exposures for the following reasons: i) some BS might have been present on the surface but not captured by OSIRIS observations because Rosetta was pointing elsewhere or because they fully sublimated between two consecutive OSIRIS sequences covering a given region; ii) in this catalog we included only the BS observed within color sequences, thus we do not consider those captured by a single filter where the spectrophotometric analysis cannot be performed; iii) we considered only the bright spots larger than 3 pixels in size; iv) in the case of clusters of icy exposure, not all the individual points (often smaller than 3 pixels in size) were counted. \\
      
For each bright feature we computed the surface, the coordinates, the spectral slope, and the minimum duration, when possible, and for a few showing negative spectral slope values we also estimated the water ice abundance using geographical mixtures of the comet dark terrain and water ice. 
The full catalog of bright spots is reported in Table~\ref{catalog}, and their distribution across the nucleus is shown in Fig.~\ref{bsmap}.

\section{Bright feature distribution and type}

\begin{figure*}
\centering
\includegraphics[width=1.0\textwidth]{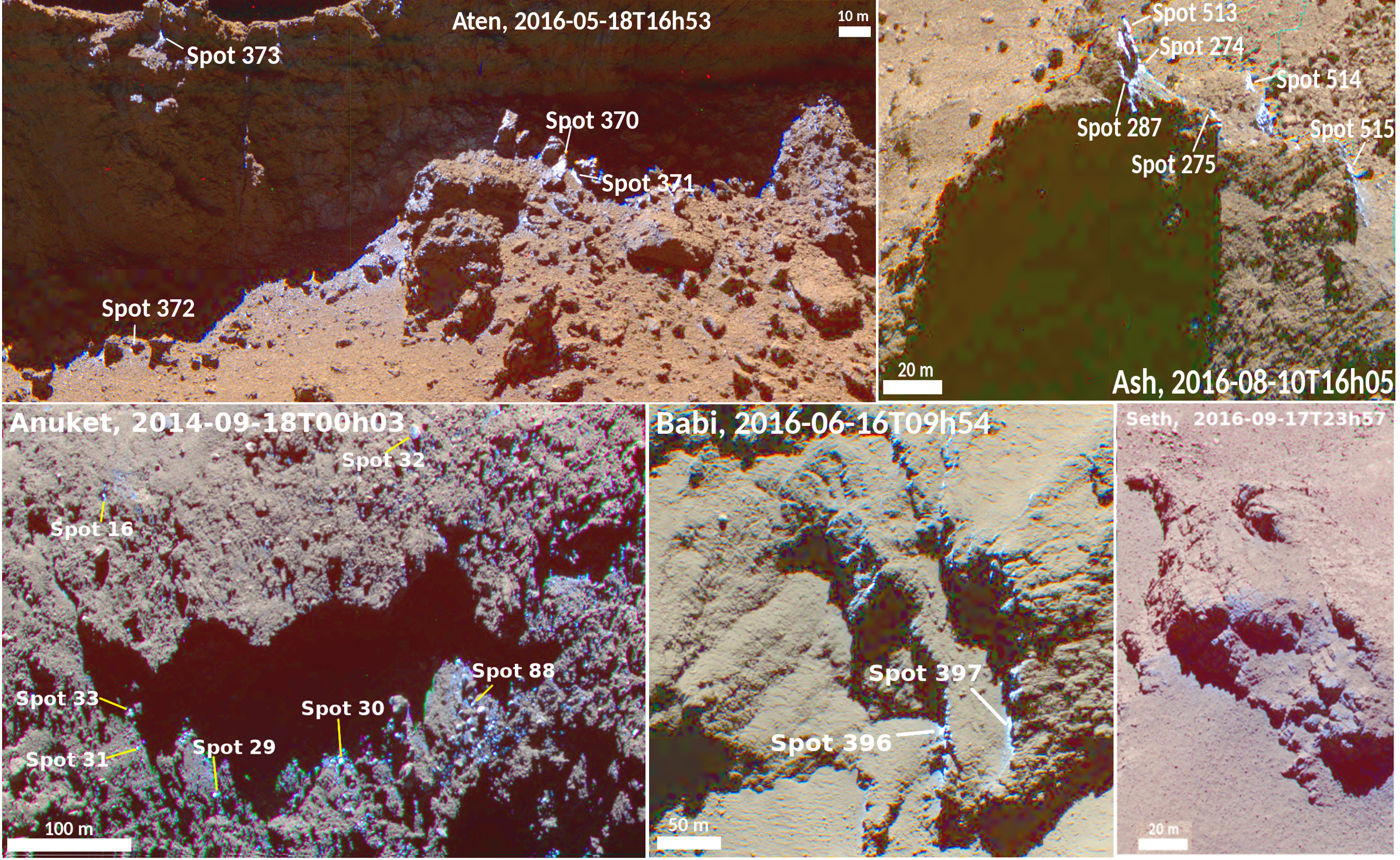}
\caption{Example of clusters of bright spots, type 4 following the  Deshapriya et al. (2018) classification scheme. The bright spot numbers correspond to those listed in Table~\ref{catalog}. Several BS show blue colors, indicating a small or negative spectral slope value (see Table~\ref{catalog}).}
\label{spots4}
\end{figure*}
\begin{table}
         \begin{center} 
         \caption{Volatile exposures types from the catalog here presented (Table~\ref{catalog}) following the  Deshapriya et al. (2018) classification scheme. }
         \label{tab_BS_type}
{\small
        \begin{tabular}{|l|c|c|c|} \hline
Feature type & number & $<area>$ & $<duration>$  \\ 
             &       & [m$^2$] & [days] \\ \hline
1 Isolated BS on smooth               & 27 & 371 & 18.1\\
terrains                                &    &     &      \\
2 Isolated BS close to    & 373 & 86 & 40.5 \\
irregular structures        &     &    &      \\
3 BS resting on boulders                      & 58 & 24.5 & 71.6\\
4 Clusters of BS                              & 145 & 23 & 35.9 \\ \hline

 \end{tabular}
}
\end{center}
\tablefoot{BS=bright spots. The average area and duration are reported for each type.}
 \end{table}

\begin{figure}
\centering
\includegraphics[width=0.45\textwidth]{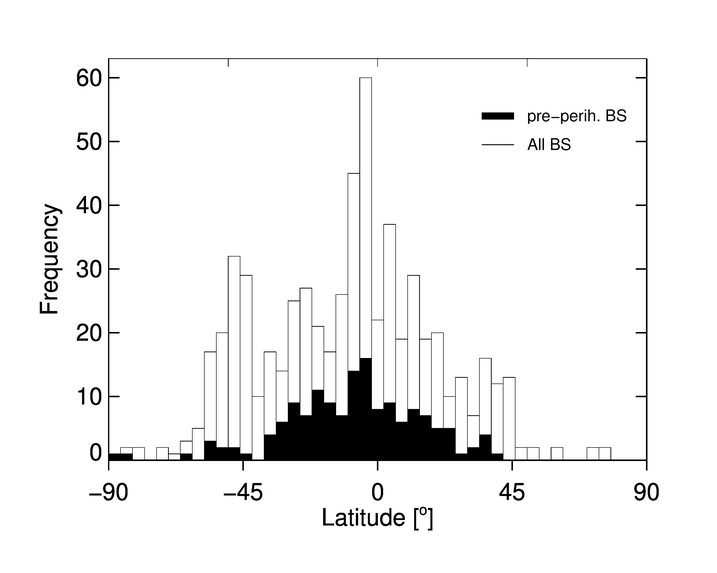}
\caption{Frequency of the bright spots vs. latitude. The histogram in black represents the pre-perihelion BS. }
\label{bs_lat}
\end{figure}

Bright patches and spots could be found isolated on the nucleus surface or grouped in clusters, usually at the bottom of cliffs. In Tables~\ref{tab_BS_type} and ~\ref{catalog} we report the BS type following the Deshapriya et al. (2018) classification scheme: type 1) isolated BS on smooth terrains; type 2) isolated BS close to irregular structures; type 3) BS on top of boulders; type 4) clusters of bright patches and BS. \\
Examples of the different types of volatiles exposures are reported in Figs.~\ref{spots1e2}, ~\ref{spots3}, and ~\ref{spots4}, while in Table~\ref{tab_BS_type} we summarize the BS identification per type. The majority of them are of type 2, and are thus identified close to irregular structures. This is quite expected because of the complex geomorphology of the comet.  \\
The largest icy exposure belongs to type 2 and was observed on Imhotep shortly after the perihelion passage, on 23 August 2015 (BS 188 in Table~\ref{catalog}). This bright patch occupied a vast surface of $\sim$ 5260 m$^2$, and was repeatedly observed for 4 hours by NAC sequences capturing that region, and was still observed one week later, even if part of it sublimated during this time lapse. It is worth mentioning that this area was brighter and spectrally bluer than the comet dark terrain, but its spectral slope was relatively high compared to other BS (around 10 \%/(100 nm)), indicating a local surface enrichment of volatiles, but highly mixed with the cometary dust. In addition, the spatial resolution was relatively low (about 6 m/px), impeding an accurate study of this BS.    \\
More than one-third of the type 2 BS are smaller than 1 m$^2$, and the average size is of 86 m$^2$, or 72 m$^2$ when excluding the largest patch previously described. For 136 out of 373 BS of type 2, we estimated their minimum duration (i.e.,  the time between the first and last sequences capturing a BS), and its average value is 41 days. \\
Smooth terrains (type 1, Fig.~\ref{spots1e2}) host few BS, but they tend to be larger ($\sim$ 370 m$^2$) than the isolated ones observed close to irregular structures or on boulders, and their average duration is the shortest  (about 18 days, Table~\ref{tab_BS_type}). This may be attributed to mixing processes with the surrounding dust, and/or to a longer illumination time and/or intensity compared to rough terrains where mutual shadows favor a longer ice survival. \\
Conversely, BS on boulders (type 3, Fig.~\ref{spots3}) are found to be smaller in size but with the longest duration (72 days). This may be associated with the presence of fractures and small cavities on boulders, which slow down the volatile sublimation. Bright spots on boulders  may also be fed by internal reservoirs of volatiles. During the second Philae touch down, the lander imprinted in a boulder, revealing a 3.5 m$^2$ bright area containing the primordial water ice embedded inside it (O'Rourke et al. 2020). They estimated a water ice fraction of 46\%  and a dust-to-ice mass ratio of 2.3 in this boulder.   \\
Clusters of bright spots (type 4, Fig.~\ref{spots4}) are located at the bases of cliffs and likely formed as a result of cliff collapses, such as the large clustered features (CFs) named CF1, CF2, and CF3 described in Oklay et al. (2017). Some of them, especially in post-perihelion images, look clearly associated with frost recondensation, like the ones in the  Ash and Babi regions shown in Fig.~\ref{spots4}. We identified 145 BS in clusters, located mainly in the Anuket, Ash, Aten, Babi, Geb, Hatmehit, and Seth regions. The individual BS in clusters are relatively small (23 m$^2$), and some of them are long-lived features; for example, the blue enriched areas in the Seth alcove (bottom right part of Fig.~\ref{spots4}) was observable for more than two years. Tiny spots (i.e., under 1 m$^2$) were frequently identified inside the Hatmehit rim in high-resolution post-perihelion images (Hoang et al. 2020), but mostly with a duration of a few minutes or 1 day, indicating the presence of frost, with the notable exception of a few BS that survived for 6--8 days. 

Volatile exposures are found at different latitudes post-perihelion. During the pre-perihelion observations they are more concentrated in the equator and mid-latitudes, between -40$^{\circ}$ and 40 $^{\circ}$ (Fig.~\ref{bs_lat}).

\begin{figure}
\centering
\includegraphics[width=0.49\textwidth]{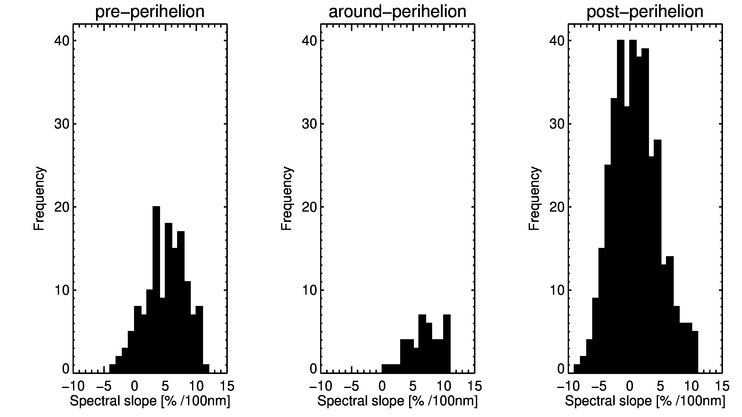}
\caption{Histograms showing the bright spot spectral slopes, evaluated in the 535--882 nm wavelength range,   pre-perihelion (August 2014-May 2015), during perihelion (June-October 2015), and post-perihelion (November 2015 - September 2016).}
\label{histoslopes}
\end{figure}



\section{Spectral slope distribution of the bright spots}

\begin{figure}
\centering
\includegraphics[width=0.49\textwidth]{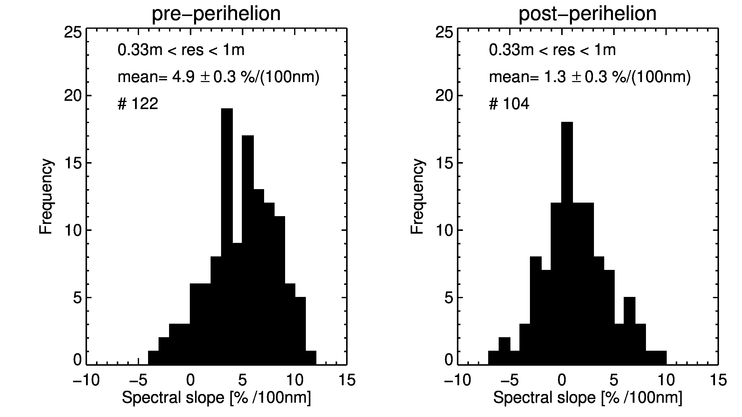}
\caption{Histograms showing the bright spots spectral slopes, evaluated in the 535-882 nm wavelength range, in images having similar high resolution (in the 0.33-1 m/px range), in the pre-perihelion (August 2014-May 2015), and post-perihelion (November 2015 - September 2016) periods.}
\label{histoslopesres}
\end{figure}

We investigated the spectral slope distribution, evaluated in the 535--882 nm range,  of the BS during the different comet orbital periods (Fig.~\ref{histoslopes}), which we defined as follows: pre-perihelion from August 2014 to the end of May 2015; during perihelion from June to October 2015; post-perihelion from November 2015 to the end of the Rosetta mission on 30 September 2016. \\  
The volatile exposures show a distinct spectral slope distribution in the post-perihelion period,  with the presence of several bright spots having negative spectral slope values, while   pre-perihelion and during perihelion   the BS spectral slope values were close to zero or moderately positive, with very few BS showing negative values (Fig.~\ref{histoslopes}). It should be noted that the observations at perihelion were acquired at relatively high distances and low spatial resolution (3-10 m/px), preventing the identification of square-meter-sized bright spots. The observing conditions explain the limited number of bright spots observed close to perihelion. 
The average spectral slope values in the 535--882 nm range of pre-perihelion, perihelion, and post-perihelion bright spots are 5.0$\pm$0.3 \%/(100 nm), 6.8$\pm$0.4 \%/(100  nm), and 0.9$\pm$0.2\%/(100 nm), respectively. To exclude that the lower spectral slope values of the BS in the post-perihelion images are related to spatial resolution effects, we investigated the BS spectral slope distribution for pre- and post-perihelion data acquired at similar high resolution,  between 0.33 m/px (the highest pre-perihelion resolution available) and 1 m/px. The histograms shown in Fig.~\ref{histoslopesres} confirm the trend observed in Fig.~\ref{histoslopes}, and thus the decrease in  the BS spectral slope in post-perihelion images. The average spectral slope values of the BS investigated at similar high resolution is of 4.9$\pm$0.3 \%/(100 nm) for the pre-perihelion BS, and 1.3$\pm$0.3 \%/(100 nm) for the post-perihelion BS.  \\

\begin{figure*}
\centering
\includegraphics[width=0.9\textwidth]{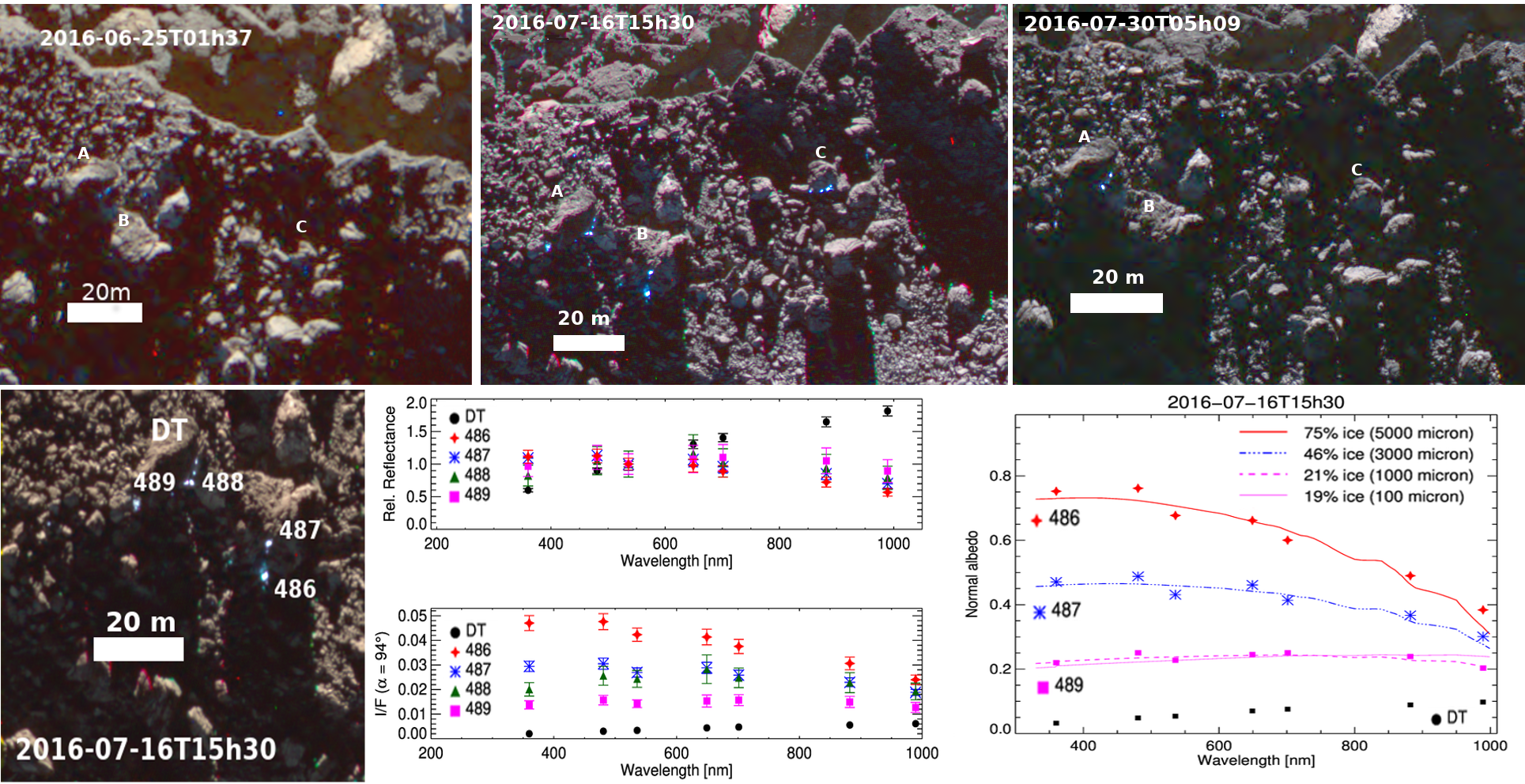}
\caption{Duration, spectrophotometry and compositional modeling of some blue bright spots in the Anhur region. Top: RGB images showing the blue BS observed in boulders named A, B, and C on 16 July 2016 (top central panel). This area is located close to the canyon-like structure of Anhur (Fornasier et al. 2017). Frosts and tiny blue spots were already present on 25 July 2016 images (top left panel) near the boulders named A and C, and a few close to boulder A were still visible in the 30 July 2016 images (top right panel). Bottom: Relative reflectance and I/F of four selected blue BS located in the aforementioned A and B boulders: BS 486 (red star, slope = -8.0\%/(100 nm)), BS 487 (blue asterisk, slope = -4.3\%/(100 nm)), BS 488 (green triangle, slope = -1.8\%/(100 nm)), and BS 489 (magenta square, slope = 4.6\%/(100 nm)) in Table~\ref{catalog}. DT indicates the cometary dark terrain. Bottom right: Linear mixing models of the cometary dark terrain and water ice with different grain sizes reproducing the  spectral behavior of the blue bright features. The estimated water ice abundance is indicated for the different models.}
\label{Anhur_BSneg}
\end{figure*}

Moreover, we   found that 57 spots  have unusually  negative slope values (i.e., below -3\%/(100 nm)) in the 535--882 nm range, which we call blue spots. All except one were observed after perihelion. The only pre-perihelion one was detected on 5 September 2014 in Imhotep (BS 42 in Table~\ref{catalog}). This region also hosts   the first blue sloped post-perihelion BS, which was detected at the end of November 2015 (BS 205 in Table~\ref{catalog}). In the same period a blue sloped BS was also observed   in Anuket (BS 199). \\
Twenty-two of these blue spots  were observed in the Anhur region (a few examples are shown in Fig.~\ref{Anhur_BSneg}), close to the canyon structure (see Fornasier et al. 2017 for the Anhur morphological description) that also hosted one of the brightest outbursts reported for comet 67P during the Rosetta observations,  called the perihelion outburst, which took place on 12 August 2015 (Fornasier et al. 2019a). Another region showing    blue spots (a total of 13) is Khonsu.  These features are located in the low bank area (i.e., a flat area between -20$^\circ$ and -30$^\circ$ latitude) defined and investigated by Hasselmann et al. (2019), where they report important morphological changes. This area was also the source of several activity events during the perihelion passage, including an outburst on 14 September 2015 (Vincent et al. 2016; Hasselmann et al. 2019).\\

\begin{figure*}
\centering
\includegraphics[width=0.9\textwidth]{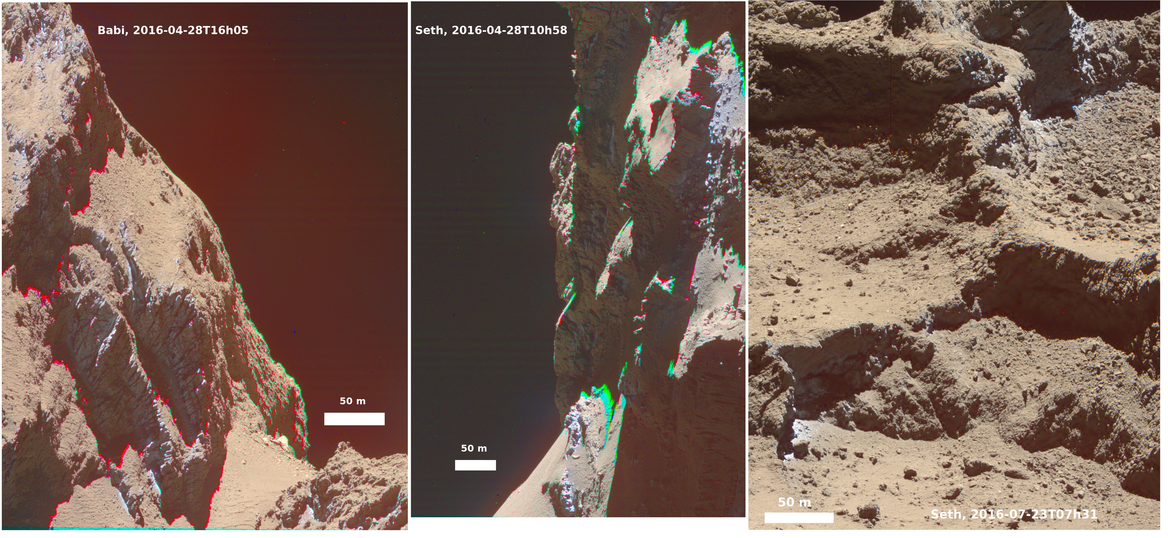}
\caption{Examples of RGB images showing frosts during the outbound orbit of comet 67P, near 3 au.}
\label{frosts}
\end{figure*}

\begin{figure*}
\centering
\includegraphics[width=0.99\textwidth]{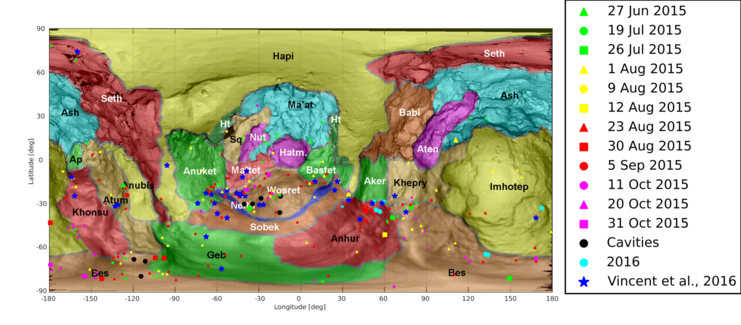}
\caption{Map with the sources of activity identified on the 67P nucleus in Vincent et al. (2016, blue asterisks), and Fornasier et al. (2019b).}
\label{mapjet}
\end{figure*}
The presence of blue spots is not restricted to the big lobe or to the southern hemisphere of the comet (where Anhur and Khonsu are located).   They  were also detected in the small lobe (e.g.,  four BS   in Wosret) and in the northern hemisphere (e.g., four BS each in the Babi and Seth regions), even if they are less frequent.  \\
The evolution of the BS spectral slope toward smaller values in post-perihelion images is also confirmed  in the analysis of the BS in individual regions, as shown in Table~\ref{tab_BS_per_region}, where the slope systematically decreases in outbound orbits compared to the pre-perihelion images for regions observed during both periods and showing at least a few   BS. It is worth noting that in Table~\ref{tab_BS_per_region} the spectral slope of the BS increased during the perihelion passage, while several studies reported a bluing of the comet colors (i.e., a decrease in the spectral slope) during the perihelion passage due to the seasonal cycle of water (Fornasier et al. 2016; Ciarniello et al. 2016, 2022). The higher spectral slope value of BS at perihelion is very likely an observational bias, due to the low spatial resolution of the data. In fact, BS are usually meter sized, thus during perihelion their spectral behavior is highly affected by the cometary red dark terrain, because BS are not resolved.\\

Most of the blue spots are located in shadowed areas, as shown in Fig.~\ref{Anhur_BSneg} or in Fornasier et al. (2021, their Fig. 10), and some others were found within frost fronts at the bottom of cliffs. 
The majority of the blue spots are only a few m$^2$ in size or even smaller, revealed thus in high-resolution images, while some other spots are a few tens of m$^2$ and the largest blue spot, with a surface of about 50 m$^2$, is located in the Babi region (BS 395 in Table~\ref{catalog}). 
The estimated lifetime of the blue spots ranges from at least 20 minutes (BS 599) to nearly two weeks for BS 489 (Fig.~\ref{Anhur_BSneg}).

The first bright blue spots we found in our analysis, and which motivated a deeper study, are located in the Anhur region and were observed in high-resolution images acquired on 16 July 2016 when the comet was 3.4 au outbound, at a resolution of 0.19 m/px (Fig.~\ref{Anhur_BSneg}). BS  486 and 487 in Table~\ref{catalog} display a sharp decrease in the reflectance after 650 nm, with spectral slope value (in the 535--882 nm range) of -8 and -4.3 \%/(100 nm), respectively. These BS are sub-meter sized, with a dimension of 0.95m$\times$0.76m for BS 486 and 0.76m$\times$0.56m for the BS 487. The decrease in reflectance is not associated with image saturation, nor to local fast sublimation between the sequence color images (that last 230 s) because the order of acquisition of the filters was not in order of increasing wavelength. The sequence order was 649-535-480-360-700-882-989 nm. Moreover, some BS survived several days, as shown in Fig.~\ref{Anhur_BSneg},
excluding fast sublimation.\\
This peculiar spectral behavior was never reported in the existing literature in pre-perihelion BS, which were usually spectrally flat in the visible range. We observed only one pre-perihelion bright spot having negative slope, located in the Imhotep region (BS 42 in Table~\ref{catalog}), very likely related to frost because of its  short lifetime (a few hours). \\
 We attempted to estimate the water ice content of the Anhur blue bright patches using linear mixing model of the cometary dark terrain and water ice with different grain sizes, using the method detailed in Fornasier et al. (2016, 2019a):
\begin{equation}
R =  p \times R_{ice} + (1-p) \times R_{DT}
 .\end{equation}
Here $R$ is the reflectance of the bright patches, R$_{ice}$ and R$_{DT}$ are respectively the reflectance of the water ice and of the cometary dark terrain,  and p is the relative surface fraction of water ice.\\
 Water ice spectra were produced using the Warren and Brandt (2008) optical constants and the Hapke radiative transfer model (Hapke 2002).  To produce the absolute reflectance of the regions of interest, we applied the Hapke (2002) photometric model correction with parameters derived from Fornasier et al. (2015). However, we neglect the disk function correction, which is set to unity, because  none of the shape models produced by the OSIRIS team has a spatial resolution high enough to correctly reproduce these tiny features.
{Considering this limitation, the estimated water ice abundance is about 20\% for spot BS 489, and 75\% for the brightest spot (BS 486). Conversely to most of the pre-perihelion BS observed with OSIRIS, which were matched by linear mixtures of the cometary dark terrain and water ice with grain sizes in the 30--100 $\mu$m range (Fornasier et al. 2016, 2021, 2019a; Oklay et al. 2017; O'Rourke et al. 2020), the ones  analyzed here cannot be reproduced by water ice with relatively small grains. Even the BS showing a positive slope (BS 489, magenta square in Fig.~\ref{Anhur_BSneg}) is better fitted by a model including water ice with large grains (1000 $\mu$m) than by the model including water ice with 100 $\mu$m grains. BS  487 is about five times brighter than the comet dark terrain, and its reflectance is reproduced by a mixture including 46 \% of water ice with large grains (3000 $\mu$m). Finally, BS 486 is 7--8 times brighter than the comet dark terrain (at 650 nm), and is dominated by water ice (75\%) with  very large grains   (5000 $\mu$m, Fig.~\ref{Anhur_BSneg}).  \\
The fact that the BS spectrophotometry is reproduced by water ice with large grains  in areal mixing with the cometary dark terrain is not unique, and was already reported for the modeling of some bright patches in comet 67P observed with VIRTIS. In the infrared the spectra of the exposures of volatiles show the water ice absorption bands, permitting a deeper analysis of the composition and allows us  to better constrain the components' physical parameters (grains size, temperature). Both Raponi et al. (2016) and Filacchione et al. (2016a) found that the compositional models reproducing the analyzed BS require the presence of both areal and intimate mixtures of the cometary dark terrain and water ice, this last  having different grain sizes:  a few tens of microns for the intimate mixture, and millimeter-sized grains for the areal mixture, as found in this paper for the tiny blue BS in Anhur (Fig.~\ref{Anhur_BSneg}). The BS analyzed by Raponi et al. (2016) and Filacchione et al. (2016a) were detected pre-perihelion; they showed a positive spectral slope in the visible range, survived for 2--4 months with variable abundance of the water ice related both to the seasonal and diurnal cycle, and were observed in different regions. Unfortunately, infrared spectra of the negative sloped BS   reported here are not available because the VIRTIS infrared channel has not been operational since May 2015. The presence of   millimeter-sized grains in some water ice exposures of comet 67P may be due to the sintering of smaller water ice grains or to the growth of ice crystals produced by vapor diffused from the colder and water ice enriched material beneath the surface (Filacchione et al. 2016a). \\
However, water ice in comets is usually characterized by having small to medium grain size   values: very fine grains ($\sim$ 1-2 $\mu$m) were used to model the frost fronts in the Hapi region (De Sanctis et al. 2015), to model the overall decrease in the 67P nucleus spectral slope with the increasing activity approaching perihelion (Filacchione et al. 2016c; Ciarniello et al. 2016), to model the material excavated by the Deep Impact impactor on 9P/Tempel 1 (Sunshine et al. 2007), and were also found in the 103P/Hartley coma (Protopapa et al. 2014) and 17P/Holmes outbursts (Yang et al. 2009); coarser grains (30-100 $\mu$m) were   used to model several BS of comet 67P (Filacchione et al. 2016a; Barucci et al. 2016; Pommerol et al. 2015; Fornasier et al. 2016; Oklay et al. 2017), and also on ice patches in comet 9P/Tempel 1 (Sunshine et al. 2006).

The high water ice abundance is not exclusive for the blue sloped BS reported in Fig.~\ref{Anhur_BSneg}. In  the literature high  water ice abundances were reported for a few other BS (see Table~\ref{catalog} for the BS numbers):  over 80\% in  BS 480 (Hoang et al. 2020), 64 - 69.5\% in   BS 476 (Fornasier et al. 2021), and $\sim$ 46\% (O'Rourke et al. 2020) in the boulder onto which Philae impacted (BS 385), exposing primordial ice. All these BS, located in the Wosret region, should be fresh exposures of volatiles. Other BS in the Anhur region also have a large water ice fraction (30-40\%), for example BS 461 (magenta symbol in Fig. 13, bottom panel, from Fornasier et al. 2019a), or the few large BS (1500 m$^2$ each)   observed pre-perihelion in the Anhur-Bes boundary (Fornasier et al. 2016, BS 143 and BS 144). Oklay et al. (2017) reported long-lived BS in the  Imhotep region, with estimated water ice content up to 48\% in the feature called IF2. 

Some of the post-perihelion blue spots may also be associated with frosts, especially those having short lifetimes. Frost was repeatedly observed in areas exiting from the cometary night  or shadows, as shown in Fig.~\ref{frosts}, and was already reported in the Anhur region (Fornasier et al. 2017, 2019a), near the final landing site Abydos (Hoang et al. 2020), and in Seth (Lucchetti et al. 2017), while it was less frequently observed pre-perihelion, with the notable exception of the Hapi region (De Sanctis et al. 2015). \\

\section{Volatile exposures and cometary activity}

Volatile exposures are directly linked to   cometary activity. They may either be sources of jets or appear as the result of cometary activity that generated self-cleaning and erosion of the nucleus, or they may be produced by morphological changes that expose the underlying volatile-rich materials. \\
Several relatively faint jets were observed directly departing from BS, as reported in Fornasier et al. (2019b, see, e.g.,  their Fig. 6), and some other BS appeared after cometary activity departing from their location or surroundings (Barucci et al. 2016; Deshapriya et al. 2016; Vincent et al. 2016; Fornasier et al. 2017). \\
Examples of self-cleaning of the comet are the two large bright patches reported by Fornasier et al. (2016, 2017) close to the Anhur-Bes regions boundary (BS 143 and 144 in Table~\ref{catalog}) that suddenly appeared in April-May 2015 in a smooth area. The ice survived exposed for about 7--10 days, and then these BS fully sublimated leaving a surface spectrally indistinguishable from the surrounding terrain. Notably, in the patch located within Anhur, the VIRTIS spectrometer detected in March 2015 for the very first time the exposure of carbon dioxide ice on a comet (Filacchione et al. 2016b). The discovery of carbon dioxide and water ices in this region demonstrates that different layers of volatiles are present within the nucleus, and points to compositional heterogeneity on large scales, on the order of tens of meters, on comet 67P  (Fornasier et al. 2016). 

Examples of ices exposed after activity and morphological changes are the following: the area rich in fresh water ice six times brighter than the surrounding dark terrain that appeared in the Aswan site  after the cliff collapse observed on 10 July 2015, generated by an outburst, and that stayed exposed until at least the end of 2015 (Pajola et al. 2017a, BS168 in Table~\ref{catalog}); a 15 $\times$ 5 m$^2$ bright patch formed inside a circular basin in Imhotep after an outburst on 3 July 2016 and that lasted for at least seven weeks (Agarwal et al. 2017, BS 485); a new scarp, 140 m long and 10 m high, formed in the Bes region, near the Anhur-Bes boundary and very close to   BS  144 (where CO$_2$ ice was detected), exposing in the talus volatiles with water ice abundance estimated at 17$\pm$2\% (Fornasier et al. 2017); a second 10 m high scarp, located in the Anhur region, formed in January 2016 and showing at its base a surface of about 500 m$^2$, brighter and bluer than the cometary dark terrain, exposing water ice for at least 5 months with estimated abundances of 26-30\% (Fornasier et al. 2019a,  BS 261); the appearance of a BS on a boulder in the Khonsu region at the beginning of January 2016, surviving for more than 6 months and associated with a cometary jet close to the perihelion passage (Deshapriya et al. 2018); again in the Khonsu region, a number of important morphological changes with dust removal up to 10 m in height and exposing several ice enriched patches (Hasselmann et al. 2019); water ice exposure in  front of large expanding structures in Imhotep (Groussin et al. 2015, Deshapriya et al. 2018), observed shortly before the perihelion passage;  the clustered features named CF2 and CF3 in Imhotep (Oklay et al. 2017), related to activity events and exposing water ice with abundances of $\sim$25\% ; the mechanical action of Philae that impacted into a boulder in the Wosret region, exposing fresh ice with abundance estimated at 46\% (O'Rourke et al. 2020).

The comparison between Fig.~\ref{bsmap}, showing the BS distribution on the nucleus, and Fig.~\ref{mapjet}, displaying the sources of jets and outburst observed close to the perihelion passage, clearly indicates a correlation between activity and water ice exposures, in particular for the Anhur, Bes, Khepry, Imhotep, Khonsu, Atum, Anubis, and Anuket regions.  As already noted in Fornasier et al. (2019b), activity events are correlated with local compositional heteregeneities (i.e., with local exposure of volatiles) and activated by  solar illumination.


\section{Size distribution of the volatiles exposures}

\begin{figure}
\centering
\includegraphics[width=0.45\textwidth]{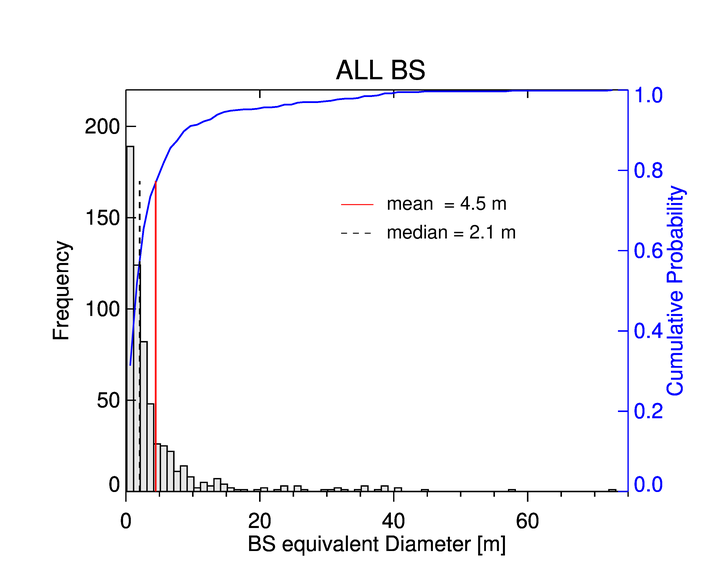}
\caption{Histogram density of the bright spot diameters identified on the 67P nucleus and reported in this study.  Only those having diameter smaller than 80 m are represented, for clarity. The cumulative probability is shown in blue.}
\label{bs-area_all}
\end{figure}

\begin{table*}
         \begin{center} 
         \caption{Bright spots per region and spectral slope values.}
         \label{tab_BS_per_region}
        \begin{tabular}{|l|c|c|c|c|c|c|c|c|} \hline
  Region         &  N$_{tot}$ & N$_{pre}$ & N$_{perih}$ & N$_{post}$& Area$_{tot}$ & Slope$_{pre}$ & Slope$_{perih}$ & Slope$_{post}$ \\ 
                 &      &      &      &      &   [m$^2$]&    [\%/ (100 nm)] &    [\%/ (100 nm)] &  [\%/ (100 nm)]       \\                \hline 
           Aker  &    1 &    1 &    0 &    0 &       3  &       7.2         &      --        &      --             \\
          Anhur  &   86 &    5 &    8 &   73 &    9512  &      3.5$\pm$2.2 &      4.8$\pm$0.9  &     -0.7$\pm$0.5   \\
         Anubis  &    5 &    0 &    4 &    1 &     701  &      --          &      7.5$\pm$0.6  &      8.1           \\
         Anuket  &   30 &   24 &    1 &    5 &    1573  &      3.4$\pm$0.7  &      9.2  &          1.1$\pm$1.3   \\
           Apis  &    0 &    0 &    0 &    0 &       0  &         --          &      --        &     --           \\
            Ash  &   63 &   30 &    0 &   33 &     466  &      5.5$\pm$0.5  &       &   2.7$\pm$0.6   \\
           Aten  &   31 &    3 &    0 &   28 &     741  &      6.5$\pm$1.8  &     -- &   1.2$\pm$0.4   \\
           Atum  &    7 &    2 &    0 &    5 &     166  &      0.2          &            --        &     1.1$\pm$0.5   \\
           Babi  &   32 &    5 &    0 &   27 &     384  &      5.5$\pm$1.5  &      --        &     0.2$\pm$0.5   \\
         Bastet  &    7 &    3 &    0 &    4 &      30  &      6.1$\pm$0.6  &      --        &     -0.8$\pm$1.2   \\
            Bes  &   25 &    4 &    6 &   15 &    8158  &      3.4$\pm$1.4  &      8.7$\pm$1.1  &      2.1$\pm$0.8         \\
            Geb  &    4 &    4 &    0 &    0 &      86  &      8.1$\pm$1.2  &      --        &      --             \\
           Hapi  &   12 &    2 &    2 &    8 &     399  &      7.0          &      --        &  1.1$\pm$1.2      \\
         Hathor  &    4 &    0 &    0 &    4 &      81  &      --           &      --        &      0.3$\pm$1.3          \\
       Hatmehit  &   30 &    1 &    1 &   28 &     466  &      7.3           &      8.8  &      2.3$\pm$0.7      \\
        Imhotep  &   80 &   38 &   13 &   29 &   15714  &      4.8$\pm$0.5  &      6.4$\pm$0.8  &      1.9$\pm$0.8         \\
         Khepry  &    9 &    6 &    1 &    2 &     202  &      3.3$\pm$1.9  &      3.9  &     -1.1$\pm$4.1         \\
         Khonsu  &   61 &   10 &    5 &   46 &    5472  &      6.4$\pm$1.0  &      7.0$\pm$1.1  &      1.1$\pm$0.7         \\
          Ma'at  &   14 &    2 &    0 &   12 &     207  &      6.3$\pm$0.3        &      --  &         1.5$\pm$0.8   \\
         Maftet  &    6 &    5 &    0 &    1 &      22  &      7.6$\pm$0.6  &      --        &    3.6              \\
          Neith  &    2 &    2 &    0 &    0 &      48  &      5.2$\pm$0.30  &      --        &      --             \\
            Nut  &    2 &    0 &    0 &    2 &       7  &      --         &      --        &      5.9$\pm$0.2    \\
         Serqet  &    5 &    0 &    1 &    4 &     214  &      --           &      5.9  &        5.2$\pm$1.6     \\
           Seth  &   54 &    0 &    1 &   53 &    1197  &      --           &     10.4  &        0.3$\pm$0.4     \\
          Sobek  &    1 &    0 &    0 &    1 &      60  &      --           &      --        &       3.7                 \\
         Wosret  &   31 &    3 &    1 &   27 &     805  &      5.6$\pm$1.4  &      9.2        &     -0.5$\pm$ 0.8          \\ \hline

 \end{tabular}
\end{center}
\tablefoot{N$_{tot}$ is the total number of bright spots per region identified, while N$_{pre}$, N$_{perih}$, and N$_{post}$ are the BS observed pre-, during-, and post-perihelion, respectively. The average spectral slope values of the bright spots are reported for the three time frames, when at least one BS is observed, with the associated errors when available (i.e., when there are at least two BS per region).}
 \end{table*}


We investigated the surface distribution of the volatile exposures. The area of the spots ranges from sub-m$^2$, for example the spots observed on the Hatmehit rim in May 2016 (Hoang et al. 2020) or in Anhur (Fig.~\ref{Anhur_BSneg}), to  a few thousand  square meters, like the two 1500 m$^2$ BS in the Anhur \& Bes regions, observed in April - May 2015 (Fornasier et al. 2016), up to  $\sim$ 5300 m$^2$ for the spot observed in Imhotep shortly after the perihelion passage (BS 188 in Table~\ref{catalog}). \\
Even if numerous bright spots are detected, the total surface of exposed water ice   reported here (Table~\ref{catalog}) is 46714 m$^2$, which is 0.1\% of the total 67P nucleus surface. This confirms that the surface of comet 67P is dominated by refractory dark terrains, while volatiles may occupy only a tiny areal fraction.\\
The majority of the volatile exposures are found in the regions of the big lobe. Their total area is about 42800 m$^2$, which correspond to 0.13\% of the large lobe surface (31.66 km$^2$, Thomas et al. 2018). The percentage of volatile exposures is six times smaller for the small lobe and the neck area, corresponding to the  Hapi and Sobek regions, with values of 0.02\% (3450 m$^2$ over 17.26 km$^2$) and 0.016\% (460 m$^2$ over 2.82 km$^2$), respectively. 

The histogram showing the frequency of the BS diameter, estimated from the BS area assuming a square shape}, is reported in Fig.~\ref{bs-area_all}. Bright spots are predominantly  small;   about one-third of them are smaller than 3 m$^2$, and 170 are smaller than 1 m$^2$. 

\begin{figure}
\centering
\includegraphics[width=0.4\textwidth]{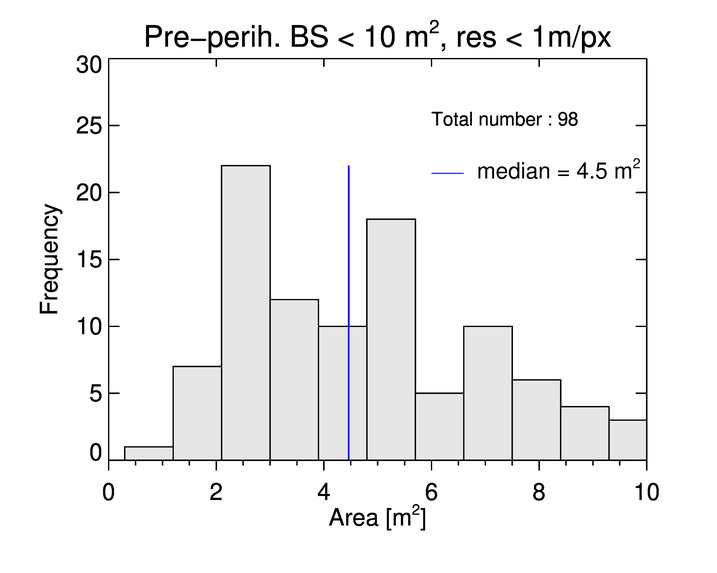}
\includegraphics[width=0.4\textwidth]{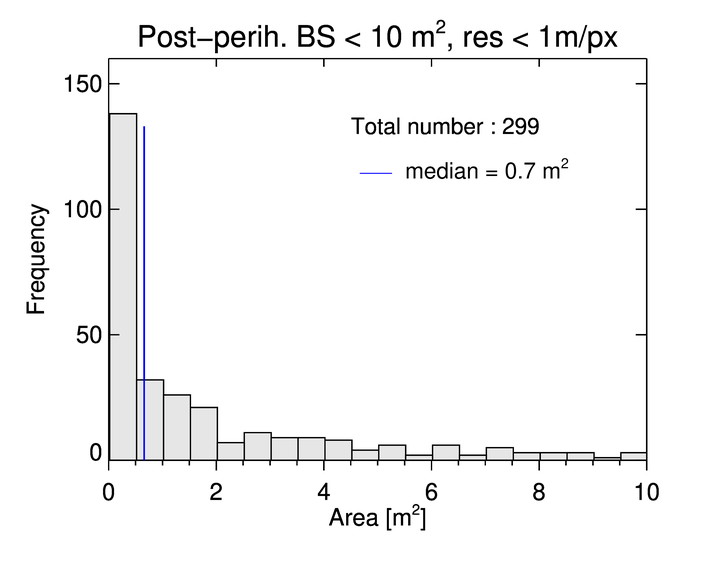}
\caption{Frequency of   bright spot areas identified on the 67P nucleus for the pre-perihelion (top) and the post-perihelion (bottom) periods, observed at spatial resolution lower than 1 m px$^{-1}$. }
\label{bs-area_prepost}
\end{figure}

\begin{figure}
\centering
\includegraphics[width=0.4\textwidth]{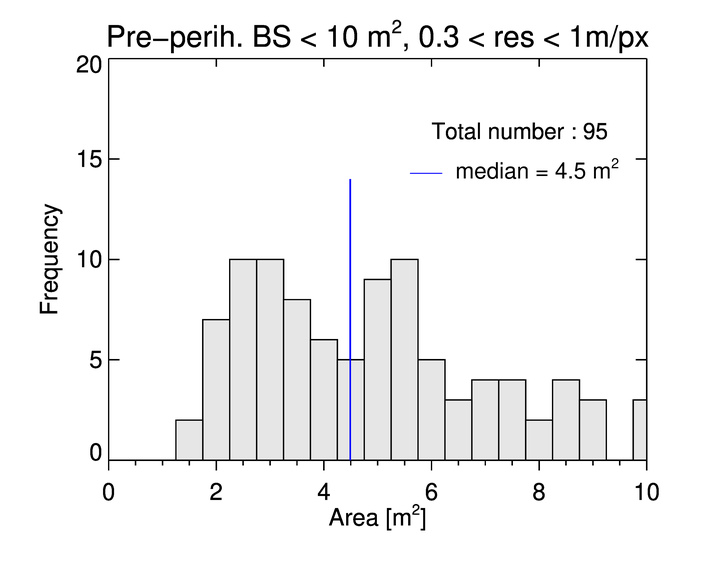}
\includegraphics[width=0.4\textwidth]{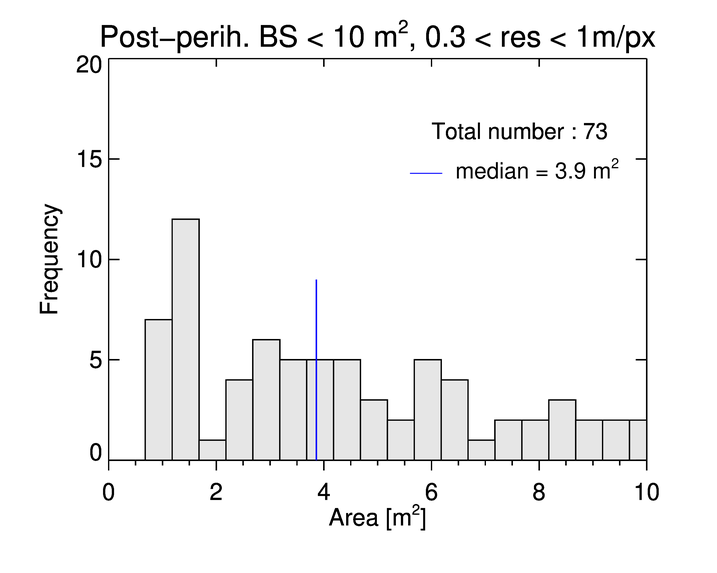}
\caption{Frequency of the bright spot areas identified on the 67P nucleus for the pre-perihelion (top) and the post-perihelion (bottom) periods, and observed at comparable high resolution (0.33--1 m/px). }
\label{bs-area_sameres}
\end{figure}

Figure~\ref{bs-area_prepost} shows the area distribution of the bright spots identified in the pre- and post-perihelion images (Table~\ref{catalog}), observed at resolutions higher than 1 m/px, and having a surface lower than 10 m$^2$; because of the low resolution, we did not consider the observations close to the perihelion passage. Even if larger bright spots were present, we chose this limit in the analysis and to compute their surface mean values because in the    pre- and post-perihelion data the great majority of these bright features have areas of a few square meters. With these criteria we count 98 pre-perihelion and 299 post-perihelion bright spots.\\
In addition to the different spectral slope distribution previously discussed, there is also a clear distinction in the areal distribution of the bright spots pre- and post-perihelion (Fig.~\ref{bs-area_prepost}). While pre-perihelion BS have a median surface of about 4.5 m$^2$, in the  post-perihelion images most of them are smaller than 1--2 m$^2$, with an average value of 0.7 m$^2$, clearly indicating that high spatial resolution is mandatory to identify ice exposures on cometary surfaces. \\
The largest number of post-perihelion BS is clearly associated with an observational bias;  approaching the end of the mission Rosetta went closer to the cometary surface achieving very high spatial resolution. More than one-third  of the BS reported in Table~\ref{catalog} (216 BS) were observed at a spatial resolution better than  0.33 m/px in post-perihelion images. When comparing the BS distribution pre- and post-perihelion observed at similar high resolution (i.e., between 0.33 m  and 1 m/px; Fig.~\ref{bs-area_sameres}), the average value of post-perihelion BS surface increases to 3.9 m$^2$, and the distribution is more compatible with the one observed pre-perihelion, even if about one-fourth of the post-perihelion BS are relatively small (< 1.5 m$^2$). 
It should be noted that small BS are still not observed at a similar resolution in pre-perihelion images. \\

We also investigated the area distribution for the different cometary regions showing more frequently exposures of volatiles. The regions having the higher number of BS (80--60) are located in the large lobe:  Anhur, Imhotep,  Ash, Khonsu, and Seth (Table~\ref{tab_BS_per_region}), while Wosret and Hatmehit, with $\sim$30 BS observed each, are the regions with more BS observed in the small lobe.

Our results on the bright spot dimensions support the  findings of Ciarniello et al. (2022) and Fulle et al. (2020), who deduced that the bright spots on comets are exposure of the water ice enriched blocks (WEBs) forming, together with the refractory matrix, cometary nuclei, and whose dominant size is on the order of 0.5--1 m. WEBs should be formed of water ice-rich pebbles mixed with drier material, and exposed to the surface when the nucleus is eroded by the cometary activity.  The fact that the majority of the bright spots are sub-meter sized is thus in agreement with these predictions and with the radar measurements the 67P comet provided by the Comet Nucleus Sounding Experiment by Radiowave Transmission (CONSERT), which indicate that the nucleus is homogeneous up to scales of a few meters (Ciarletti et al. 2017).

\section{Duration and evolution of the bright spots}
The lifetime of bright spots was estimated as the time in which a bright spot remains visible in different observing sequences, and thus it should be considered   a minimum duration (Table~\ref{catalog}).
This estimation is  biased by the observing frequency-conditions; therefore, the real lifetime is usually longer than that  reported here. 

\begin{figure}
\centering
\includegraphics[width=0.49\textwidth]{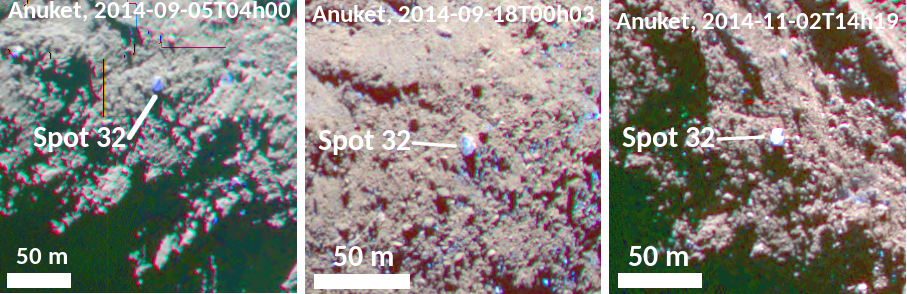}
\caption{Duration of   BS 32 in Anuket region.}
\label{ev32}
\end{figure}

\begin{figure}
\centering
\includegraphics[width=0.49\textwidth]{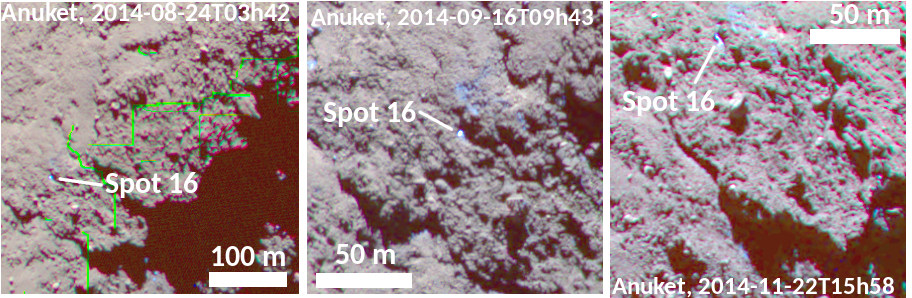}
\caption{Duration of BS 16 in Anuket region.}
\label{ev16}
\end{figure}

\begin{figure}
\centering
\includegraphics[width=0.49\textwidth]{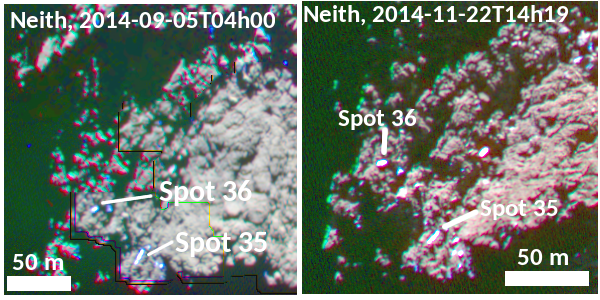}
\caption{Duration of BS 36 in Neith region.}
\label{ev36}
\end{figure}

\begin{figure}
\centering
\includegraphics[width=0.49\textwidth]{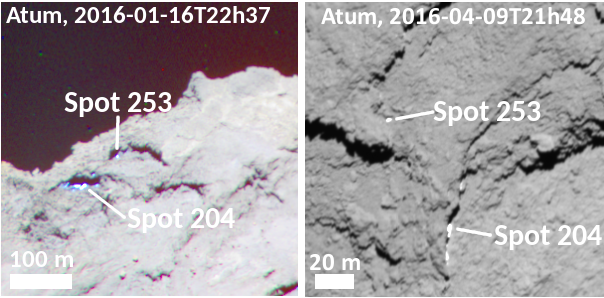}
\caption{Duration of BS 204 in Atum region.}
\label{ev204}
\end{figure}

Volatiles may survive exposed at the surface for a period varying from a few hours, in which case they are very likely frosts, to a few days and, for some of them, even to several months. A longer duration is usually found for water ice exposed after cliff collapse or the formation of new scarps, which may survive at the surface for several months, as detailed in the previous sections. \\
The evolution and water ice content of some clusters of BS was already reported in the literature, and it varies locally and with time: the CF1, CF2, and CF3 features in Imhotep contained spots with water ice fractions from 6.5\% to 24.5\% (Oklay et al.  2017); the water ice content of individual spots in the Hatmehit rim was estimated to be below 15\% in late November and early December 2015, and exceeding 50\% in late December 2015 and early 2016 (Hoang et al. 2020);   joint OSIRIS and VIRTIS data analysis of BS indicates water ice abundances ranging from 0.1 to 7.2\% (Barucci et al. 2016; Raponi et al. 2016), with seasonal variability during the BS lifetime  (on the order of 2--4 months).   \\
A few examples of BS duration are reported in Figs.~\ref{ev32}--\ref{ev204}. For pre-perihelion images, BS 32 and BS 16 in Anuket survive for at least 2 and 3 months, respectively (Figs.~\ref{ev32}, and ~\ref{ev16}), and BS 35 and BS 36 in Neith for about 2 months  (Fig.~\ref{ev36}). An example of post-perihelion duration is reported for Atum, where   BS 253 survived exposed at the surface  for at least 84 days, and BS 204 for more than 4 months (Fig.~\ref{ev204}).

\begin{figure}
\centering
\includegraphics[width=0.53\textwidth]{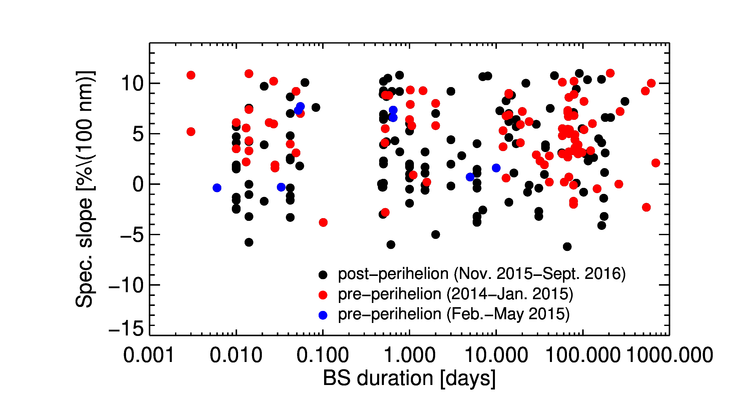}
\caption{Spectral slope value, evaluated in the 535--882 nm wavelength range, vs. BS duration.}
\label{slopeduration}
\end{figure}

Figure~\ref{slopeduration} shows the spectral slope value versus the duration for all the bright spots here reported, and for which the minimum duration was evaluated. Black points are post-perihelion BS, while red and blue points indicate the pre-perihelion observations, for the August 2014--January 2015, and February--May 2015 periods, respectively. \\
We separated the pre-perihelion observations into two datasets to investigate if any correlation exists between the blue spot color evolution and the overall evolution of the comet color and/or composition, due to the seasonal cycle of water, already observed and reported in the literature (Fornasier et al. 2016; Filacchione et al. 2016b; Ciarniello et al. 2022). The seasonal cycle of water produces on comet 67P, and very likely on comets in general, a bluing of the colors approaching perihelion, when the intense cometary activity erodes the nucleus exposing the underlying water ice enriched material. Models predict that between 20 and 70 \% of dust removed from the southern hemisphere (Keller et al. 2017; Hu et al. 2017; Fulle 2021), which is illuminated for a short time but with high intensity during the perihelion passage, falls back in the northern hemisphere mainly in decimeter-sized dehydrated aggregate, relatively poor in water ice (Keller et al. 2017). For this reason, when the cometary activity decreases, the colors of the comet go back to red values because of the dehydrated dust blanket covering the nucleus. Already at 2.2 au outbound colors were reported to be as red as in pre-perihelion observations (Fornasier et al. 2016). A substantial bluing of comet 67P surface started since February 2015, according to Ciarniello et al. (2022). These authors predict that WEBs start to be exposed mostly from this period.  Figure~\ref{slopeduration} shows that very few BS were observed in the February--May 2015 period. Most have short lifetimes of a few minutes to hours, and are very likely frosts. Instead,  the two 1500 m$^2$  bright patches observed close to the Anhur--Bes boundaries (BS 143 and BS 144 in Table~\ref{catalog}) have a spectral slope close to zero, a water ice abundance up to 30\%, and a lifetime of 1--2 weeks, and thus  are very probably  primitive WEB exposures.  Long-lived BS show variable spectral slope values, with negative values observed predominantly in the post-perihelion period, as already described in section 5. 

To further constrain whether the BS exposed after February 2015 (inbound phase) and up to the post-perihelion phase can be interpreted in terms of primordial WEB exposures, the measured BS color (spectral slope) can be directly compared with the expected value for WEBs, as computed by spectral modeling following an approach similar to Ciarniello et al. (2022), based on Hapke's theory (Hapke 2012).\footnote{In Ciarniello et al. (2022) the modeled quantity from which the color is inferred is the ``effective single scattering albedo,'' a proxy of the reflectance factor, while in the present work our computations refer to the reflectance factor itself, to be directly compared with OSIRIS observations.} WEBs are modeled as intimate mixtures of water ice and 67P dark terrain. For the dark terrain, we assume the optical properties (single scattering albedo and single particle phase function) estimated by Ciarniello et al. (2015) from VIRTIS observations (see also Ciarniello et al. 2022 for further details), while the single scattering albedo of water ice is computed from optical constants.
To constrain the color, we compute the spectral slope from the simulated spectra as a function of the ratio of the reflectance factor at 882 nm and 535 nm, following the definition given in Section 2. VIRTIS data are affected by a calibration residue in the 800--1000 nm spectral interval (Ciarniello et al. 2015; Filacchione et al. 2016a), thus the estimated reflectance of the dark terrain at those wavelengths is likely underestimated. This would result in a slight underestimation of the modeled spectral slope, as it involves the computation of the reflectance factor at 882 nm.  To bracket the correct spectral slope value, we then perform the computation also in the reduced interval 535--800 nm, where the dark terrain reflectance from VIRTIS observations is more reliable. Given the overall shape of the  67P spectrum, with a steeper slope shortward of 800 nm, this spectral slope value likely overestimates the expected value over the 535--882 nm interval.\\
In Table~\ref{speslps} we report the estimated spectral slope of different intimate mixtures of water ice and dark terrain with variable abundances. For completeness, we run simulations varying the water ice grain size (10, 100, 1000 $\mu$m), assuming this is the same as the dark material. In the model of Ciarniello et al. (2022) (hereafter WEB model), BS from WEB exposures are assumed to be characterized by a dust-to-ice mass ratio $\delta\approx2$ (O'Rourke et al. 2020), and thus their predicted spectral slope would be on the order of $\approx1.5-2.4$ $\%/(100 nm)$ (Table~\ref{speslps}). This is consistent with the average BS color during the post-perihelion phase and with the spectral slope of Anhur-Bes bright patches ($\approx0.7-1.6$ $\%/(100 nm)$) observed in April 2015, supporting the idea that such BS are exposures of original WEBs. The average BS color during the pre-perihelion phase, from observations carried out mostly before February 2015 ($\approx4.9\pm0.3$ $\%/(100 nm)$), is consistent with BS having $\delta \approx 8$, and thus likely does not represent the exposure of WEBs. The pre-perihelion BS color could also be  possibly explained by BS undergoing partial self-cleaning (see section 9).

\begin{table}
\begin{center}
\caption{Bright spot spectral slope (535--882 nm) as a function of the volumetric abundance and grain size of water ice  mixed with 67P dark terrain; delta represents the  dust-to-ice mass ratio.}
\label{speslps}
\begin{tabular}{lccccc}\\
\hline
Water ice       & $\delta$     & Spec$_{slope}$  & Spec$_{slope}$  & Spec$_{slope}$ \\
vol. ab.        &              &  [gs 10 $\mu$m] & [gs 100 $\mu$m] & [gs 1000 $\mu$m]  \\
\hline
0.9             & $\approx0.2$ & 0.56 (0.81)      & 0.50 (0.78)     & -0.055 (0.46) \\
0.8             & 0.5          & 0.83 (1.18)     & 0.79 (1.16)     & 0.40   (0.94) \\
0.5             & 2            & 1.78 (2.45)     & 1.75 (2.44)     & 1.52  (2.31) \\
0.3             & $\approx4.7$ & 2.95 (4.00)     & 2.93 (3.99)     & 2.75  (3.89) \\
0.2             & 8            & 4.11 (5.51)     & 4.10 (5.50)     & 3.95  (5.42) \\
0.1             & 18           & 6.66 (8.78)     & 6.64 (8.77)     & 6.5   (8.71) \\
\hline
\end{tabular}
\end{center}
\tablefoot{Values in parentheses are computed over the interval 535--800 nm (see text). The optical properties of the dark terrain (single scattering phase function and single particle phase function) adopted in the modeling are assumed from Ciarniello et al. (2015, 2022) and do not depend on grain size. Three different water ice grain sizes (gs) are simulated: 10, 100, and 1000 $\mu$m. For each case volumetric abundances are defined assuming that water ice and dark terrain have the same grain size. For the purpose of this work, the spectral slope is computed at a reference observation geometry with incidence angle=45$^\circ$, emergence angle=45$^\circ$, and phase angle=90$^\circ$.}
\end{table}

It is  difficult, however,  to interpret the spectral slope in terms of dust-to-ice ratio because  it depends on the spatial resolution, on the intrinsic water ice abundance in mixtures, on the water-ice grain size, and also on the spectral phase reddening. This last (i.e., the phenomenon related to the increase in colors and spectral slope values with increasing phase angle) is very important and is well characterized for the dark terrain (Fornasier et al. 2015), but has not been determined for the bright spots because of their limited phase angle coverage coupled with their relatively short visibility.


%
%
%

%
\section{Discussion}

\begin{figure*}
\centering
\includegraphics[width=0.9\textwidth]{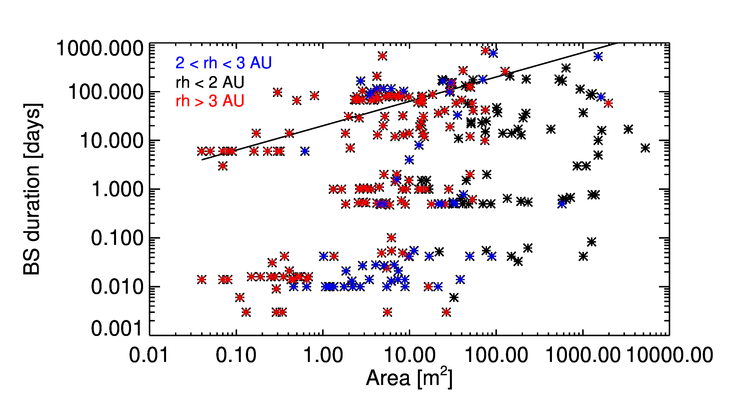}
\caption{BS duration vs. area distribution. The heliocentric distance of the 67P comet ($rh$) is indicated using symbols with different colors. The black line represents the expected lifetime of bright spots predicted by the WEB model, assuming an erosion rate of 5 cm day$^{-1}$.}
\label{slopearea}
\end{figure*}

Figure~\ref{slopearea} shows the BS duration versus their surface (Table~\ref{catalog}). It clearly demonstrates that the BS lifetimes are inconsistent with an explanation of all the observed BS in terms of diurnal or seasonal frost. The BS lifetimes do not depend on the heliocentric distance $r_h$: the largest range of lifetimes approaches six
orders of magnitudes for BS all observed at $r_h > 3$ au, thus evidencing that the main factor driving the BS lifetime
is its thickness. Since water-ice frost can be observed if its thickness is $T \ge 1 ~\mu$m, the BS with the longest lifetimes
would require $T \approx 1$ m, inconsistent with frost build-up by diurnal or seasonal thermal inversion below the nucleus
surface (De Sanctis et al. 2015; Fornasier et al. 2016) or due to the condensation on the nucleus of water vapor from the coma. \\
Bright features due to diurnal frost have a lifetime shorter than the nucleus rotation period (12.4 h, Mottola et al. 2014) and  form the lower cluster in Fig.~\ref{slopearea}, which is independent
of the BS area, because diurnal frost occurs in sunrise nucleus shadows along the terminator, which may have any area. For instance, Fornasier et al. (2016) analyze color sequences close to the perihelion passage, and find frost fronts moving with the cometary morning shadows that have a permanence time of a few minutes and a solid ice equivalent thickness of 1.5--2.7 mm.  

BS lasting more than 2 days form the upper cluster in Fig.~\ref{slopearea}, and are consistent with the WEB model (Ciarniello et al. 2022) (i.e., the proposed nucleus structural model is consistent with the observed seasonal evolution of the 67P nucleus color), which is
based on the only available activity model consistent with dust ejection (Fulle et al. 2020). WEBs are blocks of sizes ranging
from 0.5 to 1 m and composed of water-rich pebbles, embedded in a matrix of water-poor pebbles (Ciarniello et al. 2022).
Water-rich pebbles have a refractory-to-water-ice mass ratio $\delta \approx 2$ (O'Rourke et al. 2020), whereas water-poor
pebbles have $\delta \approx 50$ (Fulle 2021). Water-rich pebbles probably formed at the water-snow line of the
protoplanetary disk due to the recondensation of water vapor (Fulle 2021), so that their water ice is crystalline
and may reach millimeter sizes, consistent with the extreme blue color of some BS (Fig.~\ref{Anhur_BSneg}). Such millimeter-sized water ice is always
embedded inside the dust particles composing all the centimeter-sized pebbles structuring every comet nucleus (Blum et al. 2017).
Such a pebble structure is a necessary condition driving cometary activity (Fulle et al. 2020).

At $r_h = 3$ au, the WEB model predicts a $\delta$-independent nucleus erosion $E = 5$ cm day$^{-1}$ (Fulle et al. 2020),
best fitting the observed erosion during August--November 2014 in the Hapi deposits (Cambianica et al. 2020), which are composed of
water-poor pebbles (Fulle 2021). Therefore, the model predicts the same erosion also for all the BS lasting more than
a few days listed in Table~\ref{catalog}, which are composed of water-rich pebbles. The line in Figure~\ref{slopearea} shows the observed BS lifetimes longer than
2 days  compared to the BS lifetime predicted by the WEB model according to the computed erosion $E$, assuming that each BS is
a WEB exposed by the ongoing nucleus erosion and of thickness $T = \sqrt{A}$, where $A$ is the measured BS area (Table~\ref{catalog}).
The predicted BS lifetimes best fit the available data for $0.1 \le A \le 10$ m$^2$, with observed lifetimes longer and shorter
by up of a factor ten than predicted,  due to the fact that WEBs may have aspect ratios ranging from 0.1 to 10. WEBs of
$T < \sqrt{A}$ have  lifetimes shorter than predicted, the opposite if $T > \sqrt{A}$. On average, BS observed at $r_h < 3$ au
show a shorter lifetime (Fig.~\ref{slopearea}), as predicted by the WEB model: $E = 13$ cm day$^{-1}$ at 67P perihelion (Fulle et al. 2020).

For $A > 10$ m$^2$, most BS have  lifetimes shorter than predicted (Fig.~\ref{slopearea});  we  recall that the duration  reported here is a lower limit. This   suggests that BS of $A > 10$ m$^2$ are in fact clusters of BS of $A \le 10$ m$^2$ that appear as a single larger BS, due to the low resolution of the OSIRIS
images. Most BS of $A > 10$ m$^2$ have been observed at $r_h < 3$ au (Fig.~\ref{slopearea}), when Rosetta was farther from the
nucleus than for $r_h > 3$ au. We can conclude that the data shown in Fig.~\ref{slopearea} constrain the average WEB   cross section to
$A \le 10$ m$^2$. The cross section distribution of all BS of $A \le 10$ m$^2$ shown in Fig.~\ref{bs-area_prepost} constrains the median WEB
cross section to 0.7 m$^2$, perfectly matching the WEB model (Ciarniello et al. 2022). A total BS area covering 0.1\% of the
nucleus surface is lower than the uncertainty affecting the BS aerial fraction consistent with the seasonal evolution of the
nucleus color (Ciarniello et al. 2022). This  is  evidence that the observed BS are a few percent of the exposed WEBs driving the seasonal evolution of
the nucleus color. This may be related to the fact that most OSIRIS images have a resolution insufficient to resolve exposed sub-meter WEBs, and/or related to the criteria here adopted to identify BS (i.e.,  having  lower spectral slope values than the cometary dark terrain and higher reflectance). The average cometary dark terrain is a strong darkening agent in mixtures with ice, decreasing rapidly the BS reflectance. Thus, blue areas that are water ice enriched are often not as bright as the BS here reported (i.e., 50\%  brighter than the cometary dark terrain), and are therefore not included in this study.   

Bright spots with lifetimes between 0.5 and 2 days form another area-independent cluster, which cannot be diurnal frost of lifetime longer
than the nucleus rotation period. The most probable explanation of this cluster is the uniform fallout lasting at least the
first three months of 2016, explaining the evolution of the dust backscattering observed in the 67P coma (Bertini et al. 2019), 
and most post-perihelion nucleus reddening (Ciarniello et al. 2022). Such  uniform fallout is composed of water-poor decimeter-sized
chunks eroded by the CO$_2$-driven activity from the southern hemisphere during three months centered on the 67P perihelion
(Fulle et al. 2021). If these chunks fall back on BS where the self-cleaning is occurring (Pajola et al. 2017b), then they are
eroded into sub-chunk dust in about 2 days at the computed erosion rate of 5 cm day$^{-1}$ (Fulle et al. 2020). \\
The self-cleaning BS are never buried by the fallout
because the southern area is eroded at a perihelion rate of about 0.1 m day$^{-1}$ and because the fallout is distributed
over an area five times larger than the southern eroded area (Fulle 2021), so that the first chunk monolayer would occur in about 5 days. On the other hand, most sub-chunk dust eroded by the self-cleaning falls back on non-self-cleaning BS,
forming a dust monolayer in about 1 day (thus matching   the average cluster lifetime shown in Fig.~\ref{slopearea}) because self-cleaning and
non-self-cleaning areas are similar (Fig.~\ref{slopearea}) and because the largest falling back dust is  half the size of the parent chunks (Fulle et al. 2020). Close observations of dust deposits show that the self-cleaning areas are uniformly distributed among the non-self-cleaning areas, and that the non-self-cleaning deposits are much richer in sub-chunk dust
than the self-cleaning deposits (Pajola et al. 2017b). Non-self-cleaning deposits may be composed of chunks of $\delta > 10^4$ (Ciarniello et al. 2022), thus preventing any
water-driven erosion of the chunks themselves (Fulle et al. 2020). All of this  explains why exposed WEBs form two different clusters in Fig.~\ref{slopearea}: the area-independent
cluster which is fallout-driven and having a shorter lifetime than the area-dependent cluster,  which is erosion driven.

\section{Conclusions}

We built the most extensive catalog of exposures of volatiles on comet 67P, based on OSIRIS-NAC color images, including 603 individual entries. We investigate the type, the spectral slope, the area distribution, and their minimum duration in a homogeneous way. Our main findings are the following:\\ 

\begin{enumerate}

\item Bright spots are found isolated or in clusters, with lifetimes ranging from a few minutes--hours, in which case they are very likely frost, to several days--months, in which case they should be considered   exposure of original water ice enriched  blocks.
\item Bright spots are more often observed post-perihelion and have typical sub-meter sizes, with a median value of 0.7 m$^2$, indicating that high resolution is mandatory to observe icy exposures in comets.
\item The BS spectral slope evolved toward negative values in post-perihelion observations, indicating the presence of frost, for the short-lived ones, and of water ice with large grain sizes ($\ge$ 1000 $\mu$m) for those having longer duration. 
\item The BS lifetimes form three clusters (Fig.~\ref{slopearea}): (a) the area-independent cluster lasting less than 0.5 days, best explained
by diurnal frost; (b) the area-independent cluster persisting from 0.5 to 2 days, best explained by a seasonal fallout lasting many
months (Bertini et al. 2019, Ciarniello et al. 2022); and (c) the area-dependent cluster lasting more than 2 days, best explained by
water-driven erosion of WEBs (Fulle et al. 2020, Ciarniello et al. 2022).
\item The observed erosion of BS lasting more than 2 days and of $\delta \approx 2$ matches the one observed in Hapi's deposits
of $\delta \approx 100$ (Cambianica et al. 2020). The erosion is therefore independent of the refractory-to-ice ratio, as predicted
by the water-driven activity model (Fulle et al. 2020).
\item The observed BS lifetimes longer than 2 days are consistent with the predictions of the WEB model (Ciarniello et al. 2022).
\item The observed median of the BS cross section (0.7 m$^2$) fits that of the WEBs constrained by the seasonal evolution of the nucleus
color (Ciarniello et al. 2022).
\item The observed evolution of the BS spectral slope matches the predictions of the WEB model. After February 2015, when the
exposition of WEBs starts to make  the average nucleus color bluer (Ciarniello et al. 2022, Fornasier et al. 2016), it has a value consistent with WEBs
$\delta \approx 2$. Before this date it shows a redder color, probably due to partial self-cleaning of the few exposed WEBs. 
\item  The total integrated surface of volatile exposure is less than 50000 m$^2$, which is less than 0.1\% of the 67P nucleus surface, indicating that the top layer of cometary nucleus composition (and the whole nucleus, according to the WEB model)  is dominated by refractory material.
\item The majority of the BSs are observed in the large lobe of the comet, where they occupied 0.13\% of the surface, while the small lobe has only 0.02\% of volatile exposures. This finding supports the hypothesis of Massironi et al. (2015) on the binary structure of the comet, and the findings of El-Maarry et al. (2016) and Fornasier et al. (2021), on the fact that the small and large lobes of 67P comet have different mechanical and physical properties. Our study on volatile exposures highlights that the small lobe is pauperized in volatile abundance, at least in its upper layer, compared to the large lobe, even though  it shows a seasonal evolution of colors (Ciarniello et al. 2022), driven by the exposure of WEBs, similar to the large lobe. 
\end{enumerate}

\begin{acknowledgements}
OSIRIS was built by a consortium led by the Max-Planck-Institut f\"ur Sonnensystemforschung, Goettingen, Germany, in collaboration with CISAS, University of Padova, Italy, the Labo
ratoire d'Astrophysique de Marseille, France, the Instituto de Astrof\'isica de Andalucia,CSIC, Granada, Spain, the Scientific Support Office of the European Space Agency, Noordwijk, The Netherlands, the Instituto Nacional de T\'ecnica Aeroespacial, Madrid, Spain, the Universidad Polit\'echnica de Madrid, Spain, the Department of Physics and Astronomy of Uppsala University, Sweden, and the Institut f\"ur Datentechnik und Kommunikationsnetze der Technischen Universitat Braunschweig, Germany. \\
The support of the national funding agencies of Germany (DLR), France (CNES), Italy (ASI), Spain (MEC), Sweden (SNSB), and the ESA Technical Directorate is gratefully acknowledged.
 We thank the Rosetta Science Ground Segment at ESAC, the Rosetta Mission Operations Centre at ESOC and the Rosetta Project at ESTEC for their outstanding work enabling the science return of the Rosetta Mission. We acknowledge the financial support from the France Agence Nationale de la Recherche (program Classy, ANR-17-CE31-0004). We thank the anonymous reviewer for his/her comments and suggestions which helped us to improve this article.

\end{acknowledgements}

%
%

\begin{appendix}

\section{Supplementary material: Table}
\onecolumn
{\scriptsize
\begin{center}
\begin{longtable}{|c|c|c|c|c|c|c|c|c|c|c|l|}
\caption{List of the volatile exposure positions, types, and characteristics (slope, surface, duration) identified in this paper and in the literature} \\ \hline\hline
\label{catalog}
\textbf{BS \#} & \textbf{Start Date} & \textbf{Type} & \textbf{Selected date} & \textbf{Lon ($^\circ$)} & \textbf{Lat ($^\circ$)} & \textbf{Region} & \textbf{Res (m/px)} & \textbf{Area (m$^2$)}& \textbf{Slope} & \textbf{Duration} & \textbf{References} \\ \hline
\endfirsthead
\caption{continued.}\\
\hline\hline
\textbf{BS \#} & \textbf{Start Date} & \textbf{Type} & \textbf{Selected date} & \textbf{Lon ($^\circ$)} & \textbf{Lat ($^\circ$)} & \textbf{Region} & \textbf{Res (m/px)} & \textbf{Area (m$^2$)} & \textbf{Slope} & \textbf{Duration} & \textbf{References} \\ \hline
\hline
\endhead
\hline
\endfoot
1 & 2014-08-03T05h21 & 4 & 2014-08-03T18h39 & -178.89 & -4.49 & Imhotep & 5.03 & 1973.03 & 10.1 & 58 days & P2015, O2017 \\ \hline
2 & 2014-08-03T20h39 & 2 & 2014-08-03T20h39 & 64.75 & -48.60 & Anhur & 4.93 & 194.22 & 10.1 & -- &  \\ \hline
3 & 2014-08-05T23h19 & 3 & 2014-08-05T23h19 & -3.7 & -7.8 & Wosret & 2.2 & 50.6 & 3.31 & 123 days & D2018 \\ \hline
4 & 2014-08-06T00h19 & 3 & 2014-08-06T00h19 & -3.3 & 14.2 & Ma'at & 2.2 & 48.8 & 6 & 129 days & P2015, D2018 \\ \hline
5 & 2014-08-06T02h43 & 4 & 2014-09-05T02h35 & -54.63 & -0.58 & Anuket & 0.81 & 54.14 & 2.8 & 41 days & P2015, D2018 \\ \hline
6 & 2014-08-06T02h43 & 4 & 2014-09-05T02h35 & -56.15 & 0.42 & Anuket & 0.81 & 25.44 & 0.2 & 41 days & P2015, D2018 \\ \hline
7 & 2014-08-06T06h19 & 4 & 2014-09-03T03h44 & 115.99 & 14.60 & Ash & 1.24 & 41.45 & 7.2 & 269 days & O2017, Fi2016 \\ \hline
8 & 2014-08-14T08h43 & 3 & 2014-08-14T08h43 & -127.7 & -26.7 & Atum & 1.76 & 31 & -- & 154 days & P2015, D2018 \\
\hline
\end{longtable}
\tablefoot{BS \# is the bright spot number  assigned here; Start date is the first time a bright spot was identified in the OSIRIS color sequences;  Type represents the feature type according to the  Deshapriya et al. (2018) classification scheme.  Selected date is  the date relative to the analysis of a given bright spot to determine its surface and spectral slope; Lon, Lat, and Region are the longitude, latitude, and 67P comet region name where a bright spot is found; Res corresponds to the resolution of the images acquired in the selected date; Area and slope are the BS surface and its spectral slope in the 535-882 nm range (evaluated in the selected date); Duration is the lifetime of bright spots when it was possible to estimate it. The majority of the BS were analyzed in the paper, but some were already presented in the literature and are referenced as follows: P2015: Pommerol et al. (2015); O2017: Oklay et al. (2017); D2018: Deshapriya et al. (2018); Fi2016: Filacchionet al. (2016a); B2016: Barucci et al. (2016); H2019: Hasselmann et al. (2019); P2017: Pajola et al. (2017a); D2016: Deshapriya et al. (2016); H2020: Hoang et al. (2020); F2021: Fornasier et al. (2021); F2016: Fornasier et al. (2016); F2017: Fornasier et al. (2017); F2019: Fornasier et al. (2019a);  O2020: O'Rourke et al. (2020); A2017: Agarwal et al. (2017).}
\end{center}
}

\clearpage
\twocolumn

\end{appendix}

\newpage

\end{document}